\documentclass[aps,prd,twocolumn,floatfix,nofootinbib,nobibnotes,showpacs,superscriptaddress]{revtex4-1}

\usepackage{amsmath,amsfonts,amssymb,bm}
\usepackage{dsfont}
\usepackage{graphicx}
\usepackage{color}
\usepackage{soul}

\usepackage[caption=false]{subfig}

\usepackage[percent]{overpic}

\newlength{\panelwidth}
\newlength{\panelgap}

\newcommand{\panelplot}[2]{%
  \resizebox{\panelwidth}{!}{%
    \begin{overpic}[
      width=0.47\columnwidth,
      height=0.15\textwidth
    ]{#1}%
      \put(81,12){\textsf{\bfseries #2}}%
    \end{overpic}%
  }%
}

\usepackage[mathscr,scaled=1.15]{urwchancal}
\DeclareFontFamily{OT1}{pzc}{}
\DeclareFontShape{OT1}{pzc}{m}{it}%
{<-> s * [1.15] pzcmi7t}{}
\DeclareMathAlphabet{\mathpzc}{OT1}{pzc}{m}{it}

\definecolor{purple}{rgb}{0.5,0,0.5}
\definecolor{blue}{rgb}{0.0,0,0.9}
\usepackage[colorlinks=true, pdfstartview=FitV, linkcolor=blue, citecolor=blue, urlcolor=blue]{hyperref}

\newcounter{suppsec}

\begin{document}

\title{$\,$\\[-6ex]\hspace*{\fill}{\normalsize{\sf\emph{Preprint no}.
USTC-ICTS/PCFT-26-01}}\\[1ex]
Axial-Vector Lattice Benchmarks Reveal a Common Medium Response \\
of Meson Screening in Hot QCD
}

\author{Jie Ren}
\affiliation{Interdisciplinary Center for Theoretical Study, University of Science and Technology of China, Hefei, Anhui 230026, China}
\affiliation{Peng Huanwu Center for Fundamental Theory, Hefei, Anhui 230026, China}

\author{Chen Chen}
\email[]{chenchen1031@ustc.edu.cn}
\affiliation{Interdisciplinary Center for Theoretical Study, University of Science and Technology of China, Hefei, Anhui 230026, China}
\affiliation{Peng Huanwu Center for Fundamental Theory, Hefei, Anhui 230026, China}

\author{Fei Gao}
\email[]{fei.gao@bit.edu.cn}
\affiliation{School of Physics, Beijing Institute of Technology, Beijing 100081, China}

\author{Si-xue Qin}
\email[]{sqin@cqu.edu.cn}
\affiliation{Department of Physics and Chongqing Key Laboratory for Strongly Coupled Physics, Chongqing University, Chongqing 401331, China}

\date{\today}

\begin{abstract}
Meson screening masses trace the evolution of hadronic correlations toward quasi-free quark--antiquark screening in hot QCD.  
Combining lattice QCD (LQCD) benchmarks with a symmetry-preserving Dyson--Schwinger baseline, we identify a flavor-dependent axial-vector quasi-free onset, $x^\ast_{fg}=T^\ast_{fg}/T_c$: an operational matching scale based on the proximity of the axial-vector screening mass to its free-field reference, where $T_c$ is the pseudocritical temperature.
On the finite interval $1\le T/T_c\lesssim x^\ast_{fg}$, independent light/strange and charm-containing lattice benchmarks are organized by a common medium-response function with one flavor-sector parameter. 
One axial-vector point fixes this parameter; the remaining axial-vector data test the temperature dependence of the response, and vector screening masses validate it without vector input.
A reduced-mass interpolation then yields lattice-testable predictions for quasi-free onset scales and vector and axial-vector screening spectra in light--charm and open-bottom sectors.
\end{abstract}

\maketitle


\section{Introduction}
\label{sec1}

Quantum chromodynamics (QCD) at finite temperature describes how strongly interacting matter evolves as it is heated. At low temperature and vanishing baryon chemical potential, confinement makes color-singlet hadrons the relevant long-distance degrees of freedom\,\cite{Bazavov:2020teh}.
Lattice QCD (LQCD) calculations with physical quark masses show that the rapid change toward the hot QCD regime is a smooth crossover, rather than a singular thermodynamic phase transition \,\cite{Aoki:2006we,HotQCD:2018pds}. Because a crossover has no unique transition temperature, its characteristic location is conventionally specified by a pseudocritical temperature, $T_c$, extracted from the rapid variation or extrema of chiral and thermodynamic observables\,\cite{Aoki:2006we,Bazavov:2020teh}.
At higher temperatures, LQCD simulations and a broad body of measurements from relativistic heavy-ion collisions, interpreted together with dynamical modeling, point to the formation of a deconfined, strongly interacting regime commonly termed the quark--gluon plasma (QGP)\,\cite{Bazavov:2020teh,Harris:2023tti,Arslandok:2023utm}.

Color screening---the medium-induced weakening of long-range color interactions---is a hallmark of hot QCD\,\cite{Gross:1980br,Braaten:1995jr,Bazavov:2020teh}. Near and above the crossover, however, hadronic correlations do not disappear through a simple, abrupt dissociation process\,\cite{Bazavov:2020teh,Bala:2025ilf,Kaczmarek:2022oiu}.
Rather, they evolve under the combined influence of thermal scales and nonperturbative dynamics; in particular, the long-wavelength chromomagnetic sector remains intrinsically nonperturbative even at high temperature\,\cite{Gross:1980br,Braaten:1995jr,Bala:2025ilf}.
Identifying observables that track this evolution from hadron-like correlations to quasi-free quark--antiquark screening is therefore a central problem in hot QCD.

Spatial meson correlators are especially well suited for this purpose\,\cite{Bazavov:2020teh,Cheng:2010fe,Bazavov:2014cta,Bazavov:2019www,Petreczky:2021zmz,Bala:2025ilf,Kaczmarek:2022oiu,Aoki:2025mue,Ce:2025ihd,Laine:2003bd,DallaBrida:2021ddx}. Their large-distance behavior, $G(z,T)\sim \exp[-m_{\rm scr}(T)z]$, defines the screening mass $m_{\rm scr}(T)$. In the zero-temperature limit, this quantity reduces to the corresponding meson pole mass, whereas at high temperature it approaches the screening scale of weakly correlated quark--antiquark pairs. Screening masses therefore probe thermal dissociation, chiral-symmetry restoration, spin-dependent interactions, and the approach to perturbative or dimensionally reduced regimes\,\cite{Bazavov:2014cta,Petreczky:2021zmz,Bazavov:2019www,Aoki:2025mue,Ce:2025ihd,Laine:2003bd,DallaBrida:2021ddx}.

Although screening masses are Euclidean static observables and are not measured directly in heavy-ion collisions, they provide correlation-length benchmarks in quantum-number channels relevant to QGP spectral and transport observables. 
Related non-static vector screening correlators probe infrared physics that also enters soft-dilepton and photon production rates\,\cite{Brandt:2014uda}.
For heavy flavors, the quasi-free onsets of hidden-heavy and heavy-light correlations provide complementary thermal scales for quarkonium suppression and regeneration mechanisms\,\cite{CMS:2012gvv,Matsui:1986dk,Rapp:2008tf,Thews:2000rj} and for recombination/coalescence mechanisms for open heavy-flavor hadronization\,\cite{Greco:2003vf,He:2011qa}.

LQCD has established benchmark screening masses in the light/strange sector and in selected charm-containing channels\,\cite{Bazavov:2014cta,Bazavov:2019www}, but comparable data are still missing for light-charm and open-bottom mesons. These systems require simultaneous control of light-quark chiral dynamics and heavy-quark discretization effects. For relativistic bottom quarks, controlling cutoff effects requires the dimensionless product $a m_b\lesssim1$, with $a$ the lattice spacing and $m_b$ the bottom-quark mass; this drives thermal calculations toward very fine lattices\,\cite{Petreczky:2021zmz}. Existing bottomonium screening information is therefore used below only as a reconstructed high-temperature quasi-free anchor, not as a direct precision benchmark on the same footing as the light/strange and charm-containing LQCD data.

The Dyson--Schwinger equations (DSEs) provide a continuum, symmetry-preserving framework for nonperturbative QCD dynamics, including dynamical chiral-symmetry breaking (DCSB) and confinement-related infrared structure\,\cite{Roberts:1994dr,Roberts:2000aa,Eichmann:2016yit,Fischer:2018sdj}. Finite-temperature DSE methods can in principle access hadron screening masses\,\cite{Roberts:2000aa,Fischer:2018sdj,Maris:2000ig,Blaschke:2000gd,Gao:2020hwo}, but realistic momentum-dependent calculations across many flavor sectors and channels remain technically demanding; explicit studies of screening masses with QCD-inspired interactions are therefore scarce\,\cite{Maris:2000ig}.

In this work, we use the DSE framework as a controlled, symmetry-preserving quark-core baseline tied to external QCD benchmarks. 
Such a combination is particularly useful because LQCD provides first-principles thermal benchmarks, whereas DSEs offer direct access to flavor dependence and channel-by-channel correlations.
Specifically, we solve a vector\,$\otimes$\,vector contact-interaction (CI) model\,\cite{Wang:2013wk,Chen:2024emt,Ramirez-Garrido:2025rsu,Gutierrez-Guerrero:2026hhu} and use it to isolate the medium response required to connect this baseline with LQCD screening masses. Although the CI is a Nambu--Jona-Lasinio-type low-momentum interaction, it preserves the Ward--Takahashi identities relevant for ordinary chiral restoration and parity-partner convergence\footnote{Technical details of the finite-temperature DSE framework, the CI model, thermal dressed-quark masses $M_f(T)$, and the $T=0$ spectrum are given in Secs.\,\ref{supsec1}--\ref{supsec3} of the Supplemental Material.}.

Let $x=T/T_c$. For each flavor sector $f\bar g$, $x^\ast_{fg}$ denotes an operational matching scale on the rising branch of the transverse axial-vector (AX\footnote{In this work, AX (VC) denotes the transverse mode of an axial-vector (vector) meson, and AX-L (VC-L) the corresponding longitudinal mode.}) screening mass. We refer to this scale as the quasi-free onset, using proximity to the corresponding free-field screening mass as the defining criterion. The matching prescription and its resolution with the available lattice data are specified below.

After normalization to the corresponding free-field limits, the available LQCD-constrained AX benchmarks from independent light/strange and charm-containing calculations are organized on $1\le x\lesssim x^\ast_{fg}$ by a common functional form $\mathfrak R(x;\beta_{fg})$, with one flavor-sector parameter $\beta_{fg}$. A single AX value at $x^\ast_{fg}$ fixes $\beta_{fg}$; the remaining AX points test the temperature dependence of $\mathfrak R$, while the VC screening masses provide a non-input validation. 
We also test this response without readjustment in the ordinary CI pseudoscalar (PS)--scalar (SC) chiral-partner sector, where its extension beyond $J=1$ remains exploratory.

Finally, a Monte-Carlo Schlessinger point method (SPM)\,\cite{Schlessinger:1966zz,Schlessinger:1968vsk,Tripolt:2016cya,Chen:2018nsg,Cui:2020rmu,Cui:2021vgm} interpolation in the reduced mass $M_R$ is used to estimate $x^\ast_{fg}$ in sectors without direct LQCD benchmarks, yielding lattice-testable predictions for light-charm and bottom-containing screening spectra.



\section{CI baseline and LQCD benchmarks}
\label{sec2}

Using the thermal dressed-quark masses from the CI gap equation, we solve the finite-temperature Bethe--Salpeter equations (BSEs) for the meson screening spectrum. 
Although the CI kernel is independent of the exchanged momentum, meson correlators retain dependence on the total spatial momentum through the quark propagators. Their analytic structure determines the spatial screening scales\,\cite{Wang:2013wk}.
The $T=0$ setup provides a semiquantitative description of the known five-flavor ground-state meson spectrum, with particularly good agreement in the heavy-flavor sector; see Sec.\,\ref{supsec3} of the Supplemental Material. This establishes the CI as a controlled quark-core baseline from which deviations from LQCD screening masses can be isolated.

For the ordinary PS--SC comparison we use the minimal CI Bethe--Salpeter basis: the PS channel is represented by its leading pseudoscalar covariant and the SC channel by its single scalar covariant. This gives a clean one-covariant baseline for parity-partner comparisons\footnote{The complete PS amplitude contains an additional pseudovector covariant. Its finite-temperature effect and the reason for using the reduced one-covariant PS baseline are detailed in Sec.\,\ref{supsec4} of the Supplemental Material.}.

Before comparing with LQCD, we account for the different temperature scales and current-quark masses $m_f$ used in the lattice and CI calculations. For a setup $Y\in\{\mathrm{LQCD},\mathrm{CI}\}$, the massive free-field screening mass is
\begin{align}
\label{freeY}
m_{\rm free}^{Y,fg}(x)
&=
\sqrt{(\pi xT_c^Y)^2+(m_f^Y)^2}
+
\sqrt{(\pi xT_c^Y)^2+(m_g^Y)^2}\,.
\end{align}
For the light/strange sectors $u\bar d$, $u\bar s$, and $s\bar s$, we follow the LQCD convention\,\cite{Bazavov:2019www}
 and use the massless limit
\begin{align}
\label{freeYl}
m_{\rm free}^{Y,fg}(x)=2\pi xT_c^Y\,.
\end{align}
The lattice points are converted to the CI convention by preserving
$m_{\rm scr}/m_{\rm free}$:
\begin{align}
\label{lqcdrescale}
\widetilde m_{\rm scr}^{\,\mathrm{LQCD},fg}(x)
:=
\frac{m_{\rm scr}^{\mathrm{LQCD},fg}(x)}
     {m_{\rm free}^{\mathrm{LQCD},fg}(x)}
     m_{\rm free}^{\mathrm{CI},fg}(x)\,.
\end{align}
Thus, the comparison is made at the level of the dimensionless ratio $m_{\rm scr}/m_{\rm free}$, expressed in the CI convention.

The uncorrected CI screening masses generally lie below the AX lattice benchmarks in the matching region. This discrepancy is consistent with the limited dynamical content of the momentum-independent CI kernel. To represent phenomenologically the net correction required to connect this baseline with the lattice benchmarks, we introduce a multiplicative response with a single flavor-dependent parameter. For $1\le x\lesssim x^\ast_{fg}$, the corrected screening mass in channel $X$ and flavor sector $f\bar g$ is defined by
\begin{align}
\label{corrmass}
m_{\mathrm{scr},X}^{\mathrm{corr},fg}(x)
=
\mathfrak R(x;\beta_{fg})\,
m_{\mathrm{scr},X}^{\mathrm{CI},fg}(x)\,,
\end{align}
with
\begin{align}
\label{corr}
\mathfrak R(x;\beta_{fg})
=
\frac{\beta_{fg}(x-1)}
{1-\exp[-\beta_{fg}(x-1)]}\,.
\end{align}
Here $X\in\{\mathrm{PS},\mathrm{SC},\mathrm{VC},\mathrm{AX}\}$. The value $\mathfrak R(1;\beta_{fg})=1$ is defined by continuity, preserving the CI baseline at $T_c$. For $\beta_{fg}>0$, the response is positive, smooth, and monotonically increasing. Eq.\,\eqref{corr} is a phenomenological parametrization selected for these properties and its simplicity; they do not uniquely determine its functional form. We adopt this form throughout the present study and leave a systematic comparison with alternative responses for future work.

\begin{figure*}[t]
\centering
\setlength{\panelwidth}{0.32\linewidth}
\setlength{\panelgap}{%
  \dimexpr(\linewidth-3\panelwidth)/2\relax
}
\makebox[\linewidth][c]{%
  \panelplot{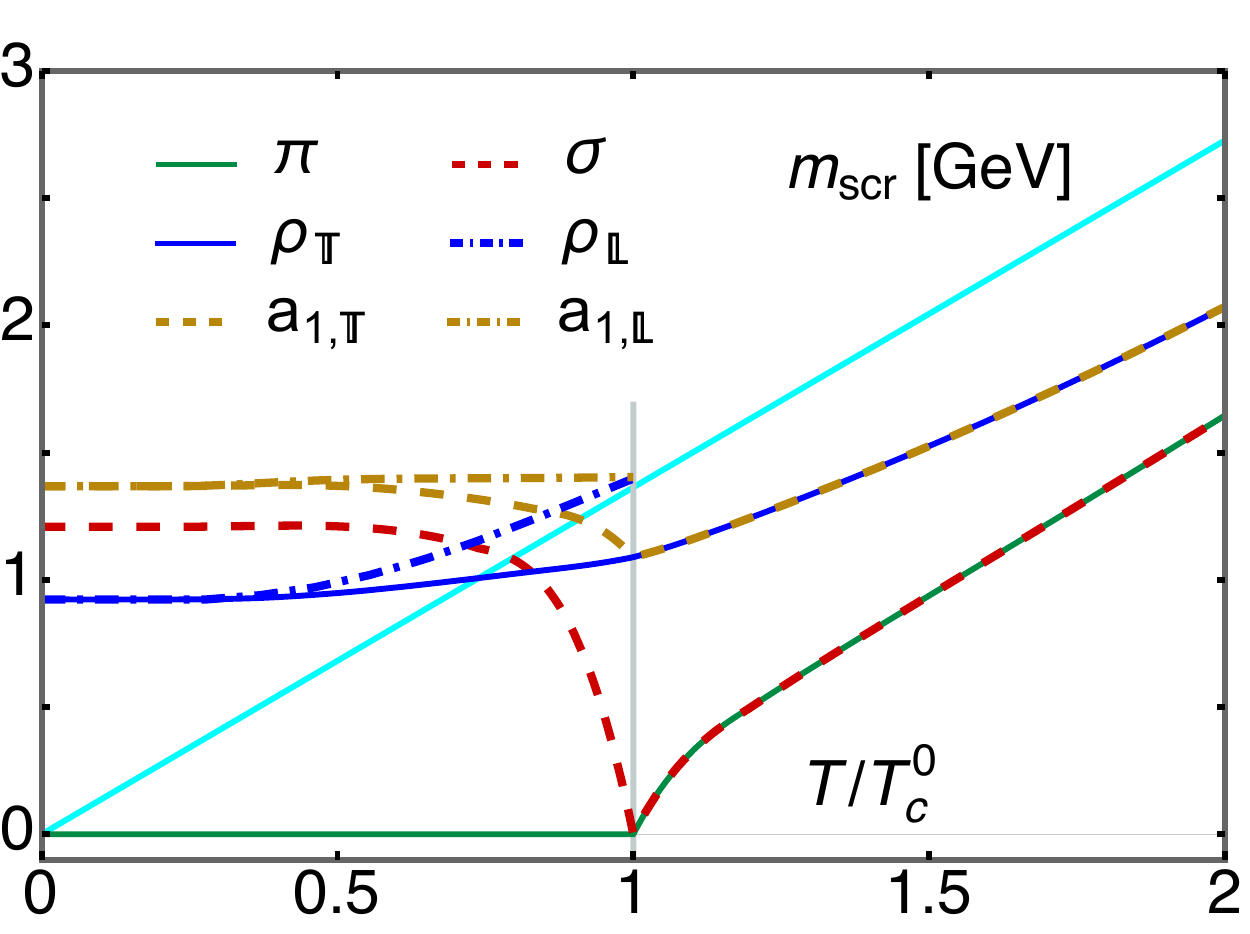}{A}%
  \hspace{\panelgap}%
  \panelplot{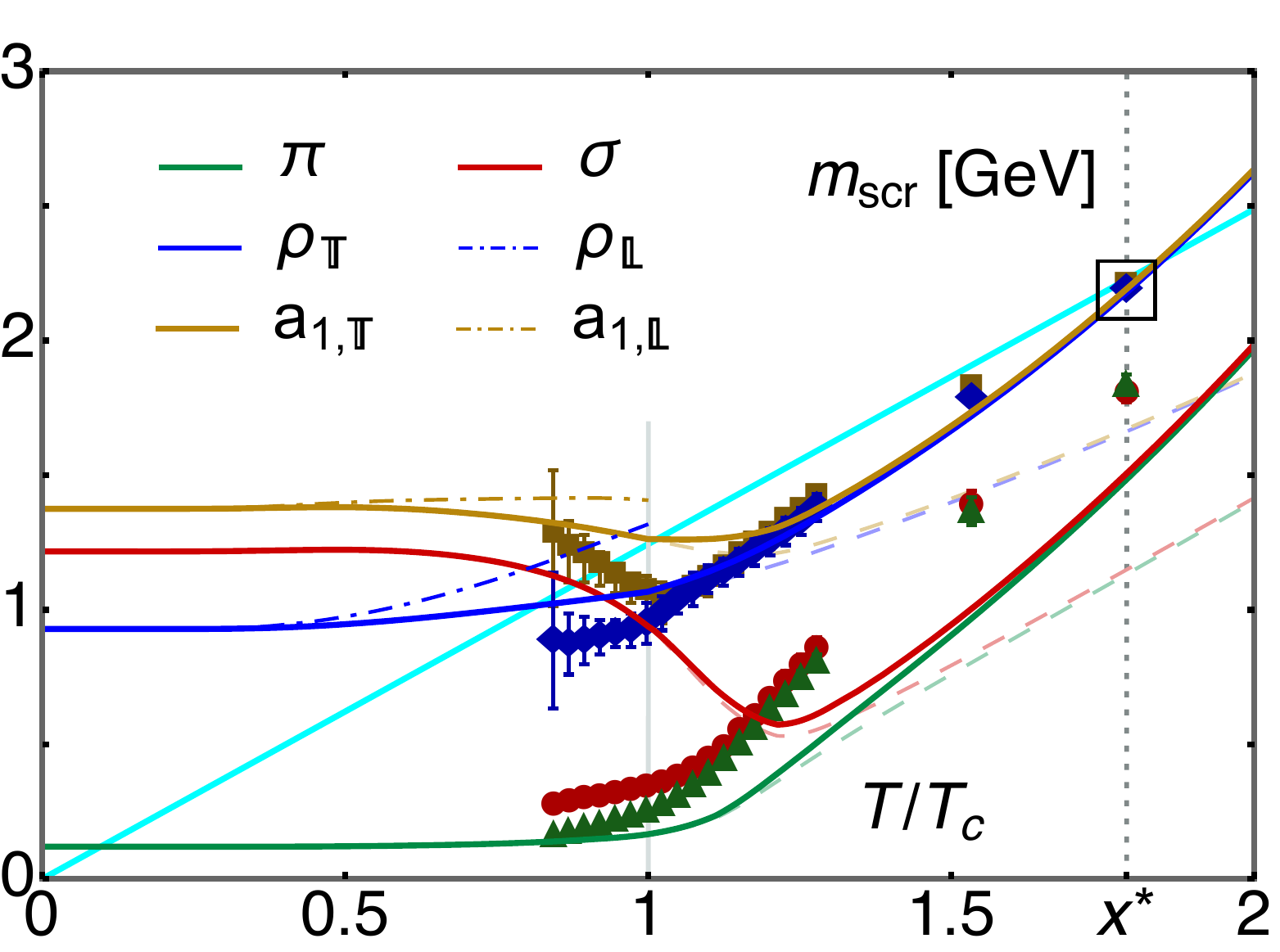}{B}%
  \hspace{\panelgap}%
  \panelplot{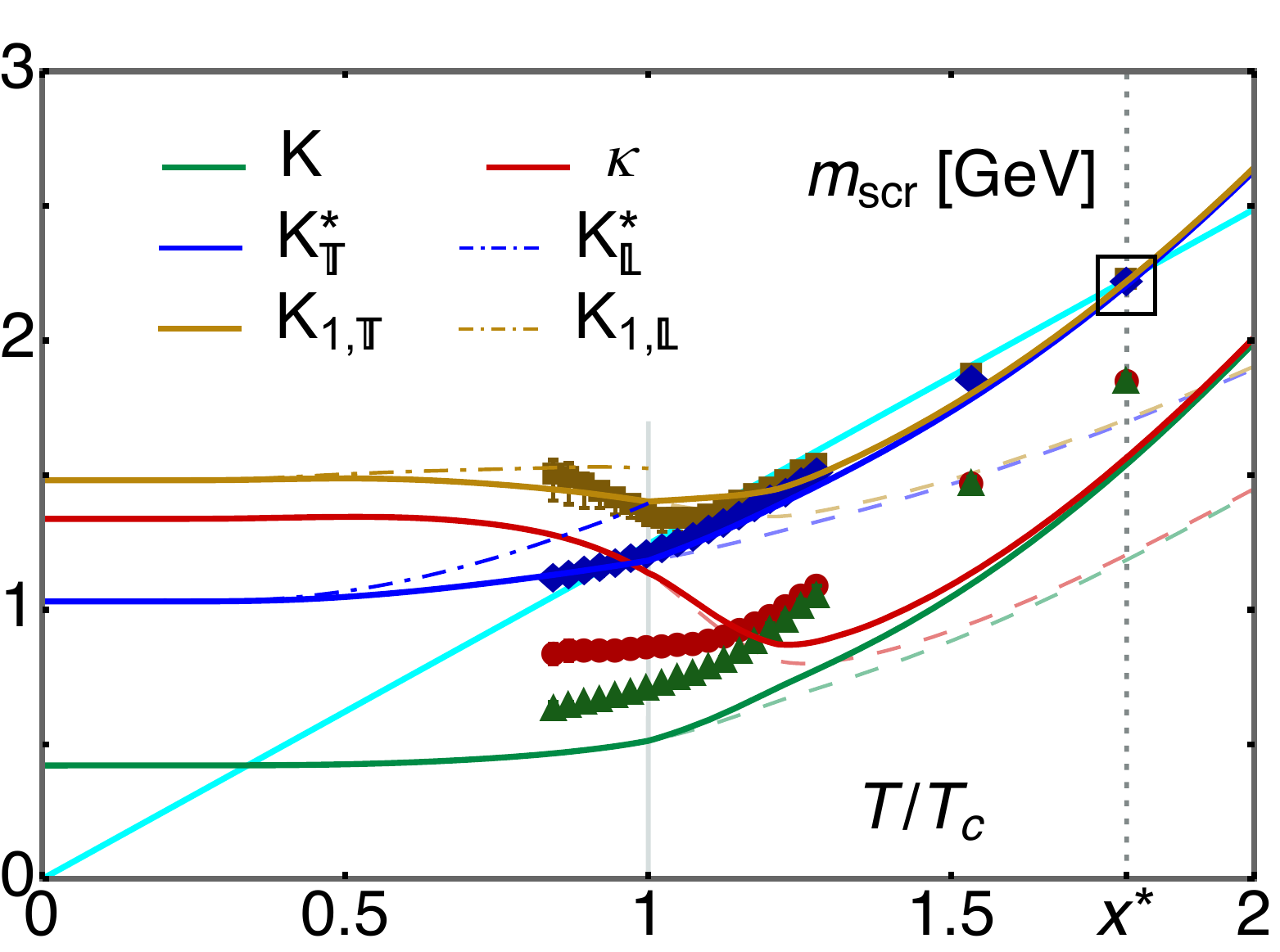}{C}%
}
\par\vspace{0.4em}
\makebox[\linewidth][c]{%
  \panelplot{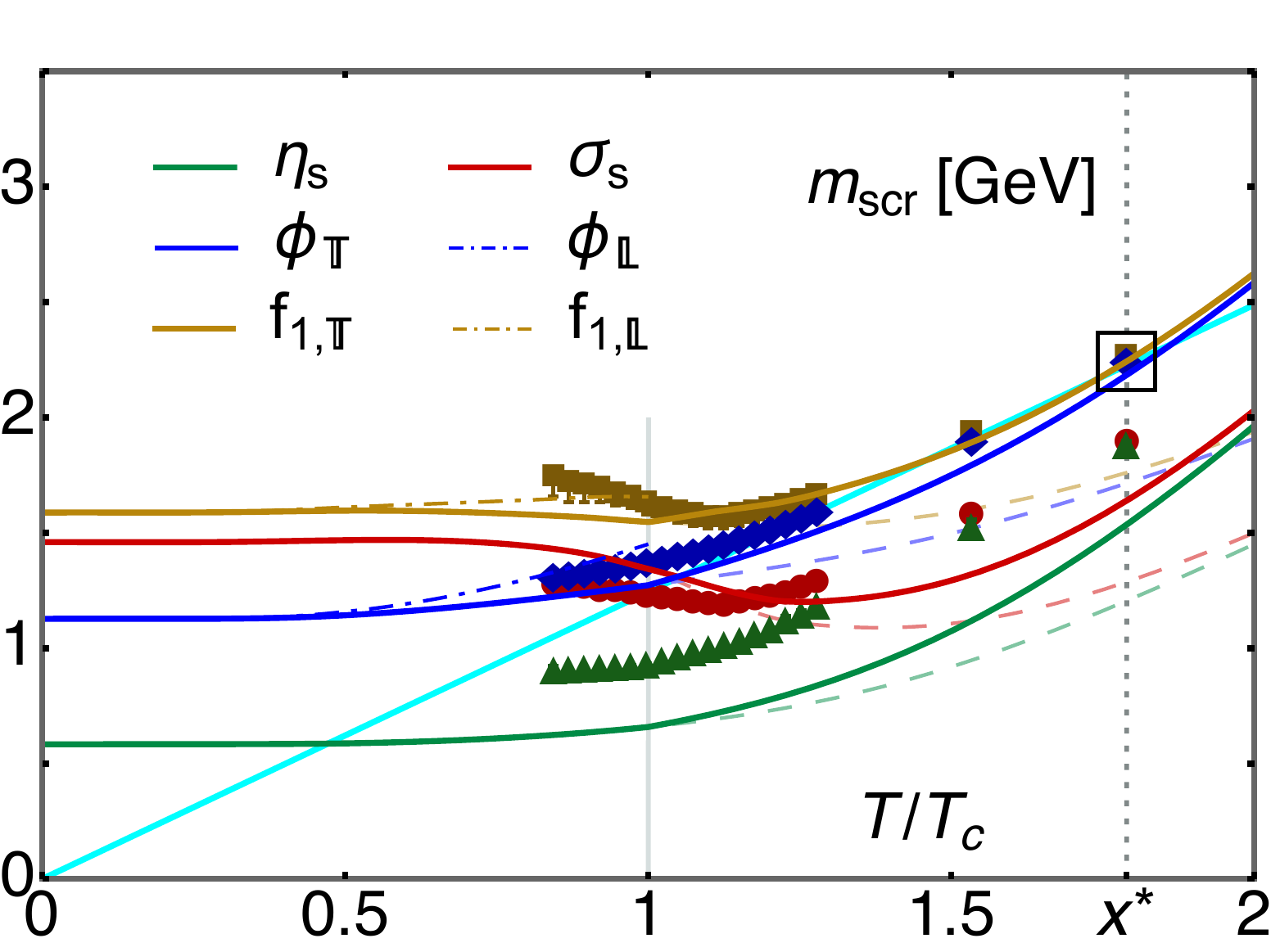}{D}%
  \hspace{\panelgap}%
  \panelplot{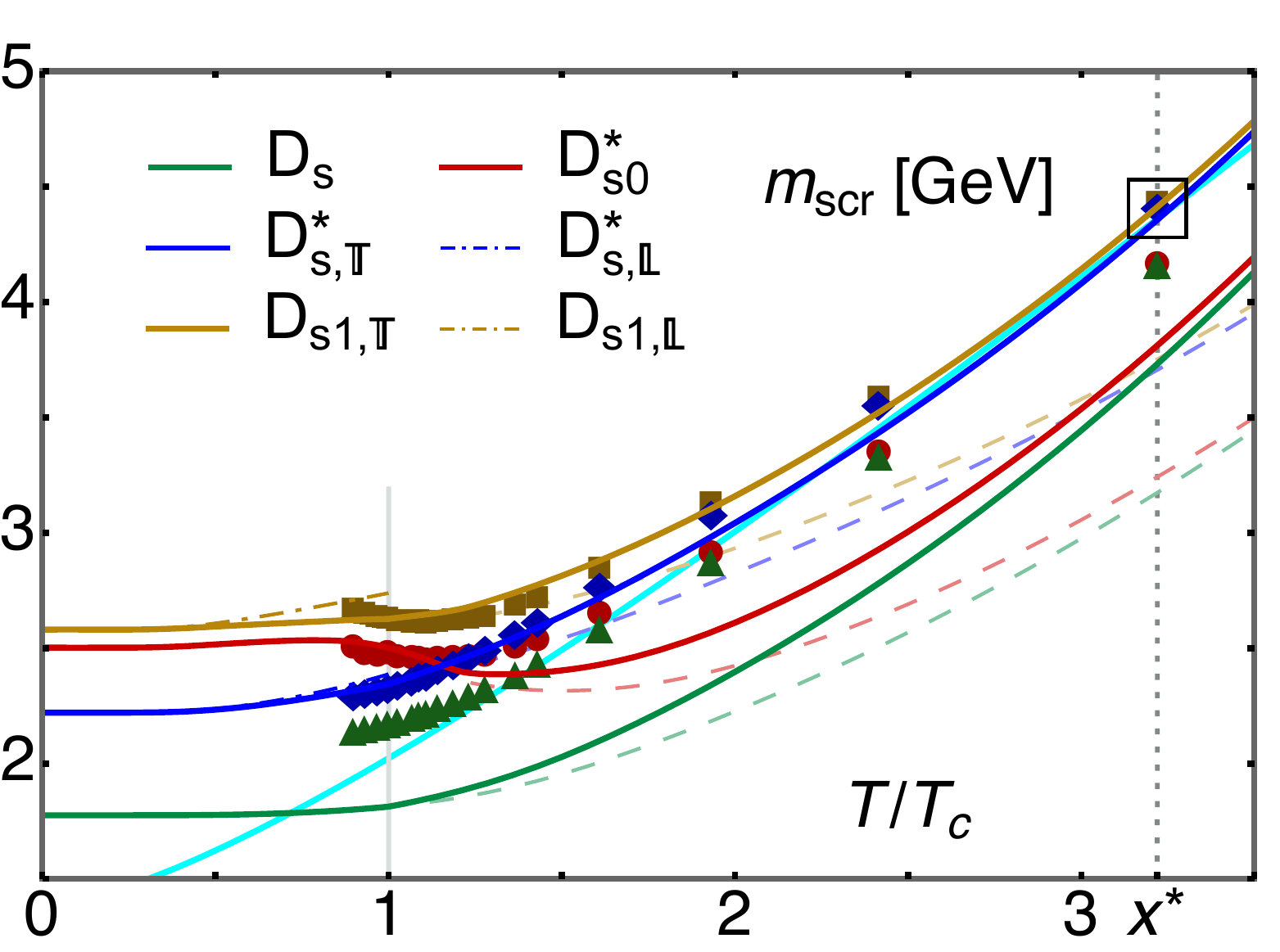}{E}%
  \hspace{\panelgap}%
  \panelplot{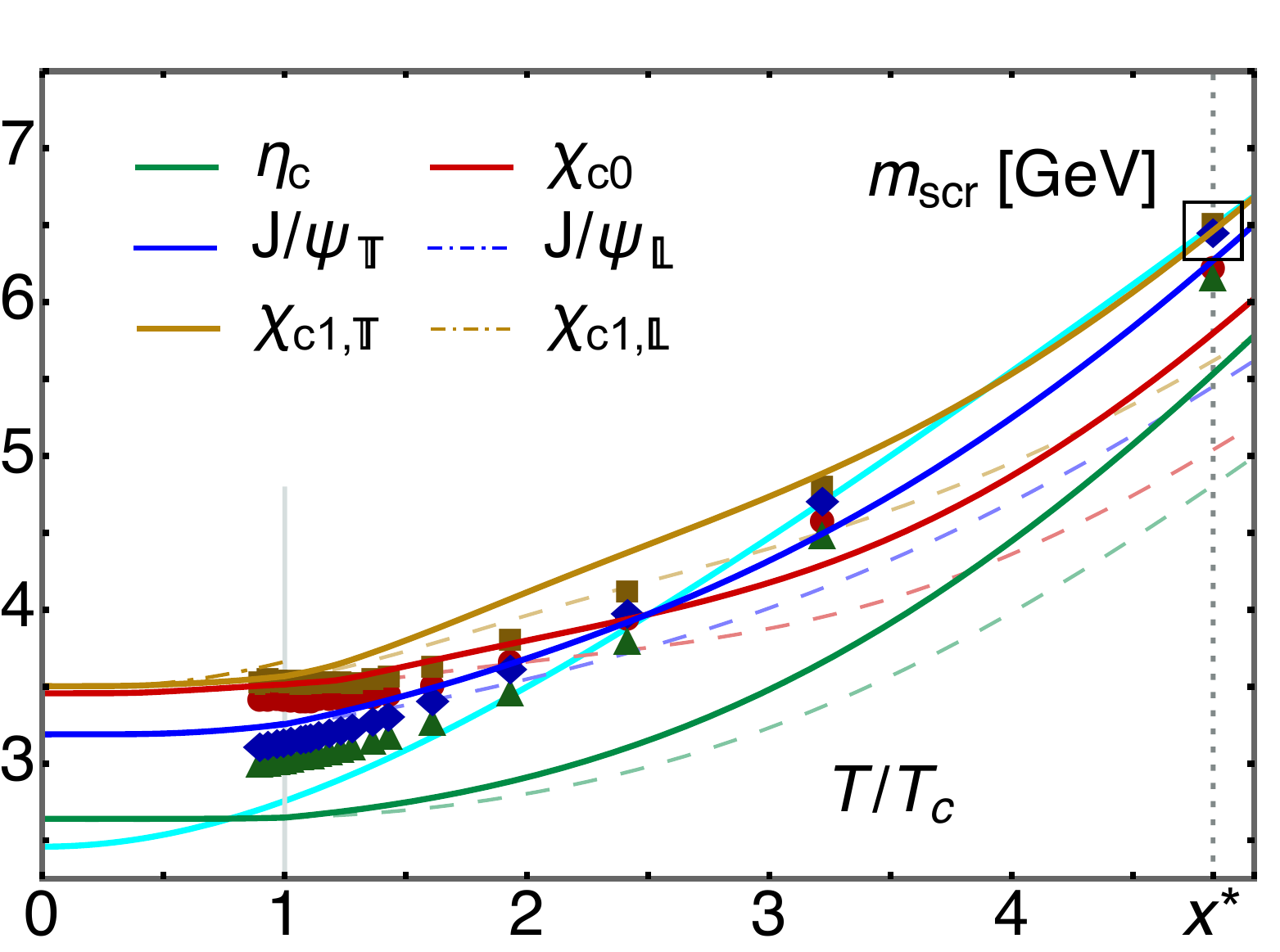}{F}%
}
\caption{\label{fig:mscr1}
CI screening masses and LQCD benchmarks in selected flavor sectors.
{\sf Panel A} shows the chiral-limit $u\bar d$ sector; {\sf panels B}--{\sf F} show physical $u\bar d$, $u\bar s$, $s\bar s$, $s\bar c$, and $c\bar c$ sectors. Curve identities are given in the panels. For $J=1$ states, $\mathds T$ and $\mathds L$ denote transverse and longitudinal modes; VC-L and AX-L are dot-dashed and have stable CI solutions only for $T<T_c$ ($T<T_c^0$ in {\sf panel A}). In {\sf panels B}--{\sf F}, the solid curves are the corrected CI results, whereas lighter dashed curves show the corresponding uncorrected CI baselines, and the vertical dotted lines indicate $x^\ast_{fg}$, the adopted AX matching point listed in Table\,\ref{tablebv}.
The finite-interval response $\mathfrak R$ is calibrated and tested on $1\le x\lesssim x^\ast_{fg}$; any continuation beyond $x^\ast_{fg}$, where shown, is only a guide to the quasi-free branch. Colored symbols are rescaled LQCD results, Eq.\,\eqref{lqcdrescale}, with colors matched to the corresponding CI channels. The open black boxes enclose the AX LQCD points used to determine $\beta_{fg}$; they do not denote additional data. 
{\sf Panels B}--{\sf D} use the continuum-extrapolated HotQCD light/strange data\,\cite{Bazavov:2019www}; {\sf panels E} and {\sf F} use the independent strange-charm/charmonium data\,\cite{Bazavov:2014cta}. Cyan curves are the CI free-field limits. 
}
\end{figure*}

The correction in Eq.\,\eqref{corrmass} is defined on $1\le x\lesssim x^\ast_{fg}$. Matching endpoints are selected from the rising AX branches in the LQCD data of Refs.\,\cite{Bazavov:2014cta,Bazavov:2019www} according to their proximity to the corresponding free-field references in Eqs.\,\eqref{freeY} and \eqref{freeYl}, using the sector-specific prescriptions below. The available temperature sampling limits the resolution of these operational matching scales; denser lattice data would permit their refinement.
We quantify this proximity through the relative deviation of the lattice central value,
\begin{align}
\label{deltaAX}
\delta_{\mathrm{AX}}^{fg}
=
\frac{
\widetilde m_{\mathrm{scr},\mathrm{AX}}^{\,\mathrm{LQCD},fg}
(x^\ast_{fg})
}{
m_{\mathrm{free}}^{\mathrm{CI},fg}(x^\ast_{fg})
}-1\,.
\end{align}
The single parameter $\beta_{fg}$ is fixed solely by the AX mass at the selected endpoint:
\begin{align}
\label{betafix}
m_{\mathrm{scr},\mathrm{AX}}^{\mathrm{corr},fg}(x^\ast_{fg})
=
\widetilde m_{\mathrm{scr},\mathrm{AX}}^{\,\mathrm{LQCD},fg}
(x^\ast_{fg})\,.
\end{align}
The resulting matching points and response parameters are listed
in Table\,\ref{tablebv}.

Based on a joint assessment of the rising AX branches in Ref.\,\cite{Bazavov:2019www}, we adopt $x^\ast_{fg}=1.79$ as a common matching endpoint for the $u\bar d$, $u\bar s$, and $s\bar s$ sectors.
This corresponds to $T=280\,\mathrm{MeV}$ for $T_c^{\mathrm{LQCD}}=156.5\,\mathrm{MeV}$.
At this temperature, all three lattice central values lie within $2\%$ of the massless free-field reference, Eq.\,\eqref{freeYl}.
For the $s\bar c$ and $c\bar c$ sectors, we select the lowest-temperature available point on each rising AX branch whose deviation from the corresponding massive free-field reference satisfies $|\delta_{\mathrm{AX}}^{fg}|\lesssim1.0\%$.
Specifically, the absolute percentage deviation, rounded to one decimal place, must not exceed $1.0\%$. This prescription selects $x^\ast_{s\bar c}=3.22$ and
$x^\ast_{c\bar c}=4.83$. The $c\bar c$ endpoint also coincides with the highest-temperature AX datum tabulated in the Appendix of Ref.\,\cite{Bazavov:2014cta}.\footnote{For the other benchmark sectors, higher-temperature LQCD data are available but lie beyond the matching intervals considered here and are not shown.}

\begin{table}[t]
\caption{\label{tablebv}
AX matching endpoints and response parameters. The endpoints $x^\ast_{fg}$ follow the sector-specific prescriptions in the text, with a common endpoint for the light/strange sectors.
The deviations $\delta_{\mathrm{AX}}^{fg}$ are defined in Eq.\,\eqref{deltaAX}, and $\beta_{fg}$ is fixed solely by the AX matching condition, Eq.\,\eqref{betafix}.
}
\begin{ruledtabular}
\begin{tabular}{lccccc}
Flavor sector
& $u\bar d$ & $u\bar s$ & $s\bar s$ & $s\bar c$ & $c\bar c$ \\
\hline
$x^\ast_{fg}$
& 1.79 & 1.79 & 1.79 & 3.22 & 4.83 \\
$\delta_{\mathrm{AX}}^{fg}\,[\%]$
& $-1.72$ & $-0.93$ & $+0.72$ & $+1.02$ & $-0.45$ \\
$\beta_{fg}$
& 0.72 & 0.70 & 0.64 & 0.15 & 0.075
\end{tabular}
\end{ruledtabular}
\end{table}

All selected endpoints satisfy $|\delta_{\mathrm{AX}}^{fg}|<2\%$, as listed in Table\,\ref{tablebv}. The endpoint AX datum is used for calibration; the remaining AX data and all VC data within $1\le x\lesssim x^\ast_{fg}$ test the resulting response. The same $\beta_{fg}$ is applied to PS, SC, VC, and AX without readjustment within each flavor sector.

Fig.\,\ref{fig:mscr1} summarizes the benchmark spectra. 
In the chiral limit, the present CI model exhibits a second-order chiral transition at $T_c^0=0.215\,\mathrm{GeV}$\,\cite{Wang:2013wk,Chen:2024emt}. Panel A shows the corresponding chiral-restoration pattern: below $T_c^0$, the pion is massless, while its SC partner decreases to zero and becomes degenerate with it at $T_c^0$; the VC--AX and VC-L--AX-L pairs show the analogous degeneracy.
Above $T_c^0$, the PS--SC and VC--AX pairs remain degenerate, while no stable VC-L and AX-L solutions are found. 
The physical-mass results in panels B--F show that the uncorrected CI curves reproduce the low-temperature flavor ordering and parity-partner pattern but depart from the LQCD benchmarks and high-temperature free-field limits. This is expected: the finite-temperature calculation should contain hot QCD dynamics beyond CI, especially those governing the approach to perturbative or quasi-free screening.


Across the two lattice data sets, the corrected CI curves reproduce the available $J=1$ benchmarks within the matching intervals, apart from two localized deviations. 
The $u\bar s$, $s\bar s$, and $s\bar c$ AX curves track the LQCD data, and the corresponding VC curves agree without using VC input. In $u\bar d$, the residual offset around $T_c$ points to a limitation of the light-sector CI quark-core baseline, possibly related to meson-loop effects absent from the CI kernel\,\cite{Chen:2012qr}. In $c\bar c$, the low-to-intermediate-$x$ discrepancy reflects the limitation of the spin--orbit ansatz, Eq.\,\eqref{gsomax}, in the heavy-heavy AX channel. Outside these localized regions, the same $\mathfrak R$ captures the AX and VC trend toward the quasi-free onset.

The PS--SC sector requires a separate interpretation. In the CI baseline, PS--SC denotes ordinary chiral partners, such as $\pi$--$\sigma$, and tracks ordinary chiral restoration or, away from the chiral limit, parity-partner convergence. Available light-sector LQCD SC points instead correspond to $U_A(1)$-sensitive channels, such as the isotriplet $a_0$, and may carry additional SC-channel systematics. We therefore use LQCD PS points as physical PS benchmarks, while LQCD SC points are shown only as reference data. 
Applying the same $\beta_{fg}$ without readjustment is motivated by the shared dressed-quark propagators, but remains a phenomenological hypothesis. Channel dependence already enters through the CI Bethe--Salpeter kernels and the spin--orbit prescription, Eq.\,\eqref{gsom}; $\mathfrak R$ introduces no additional channel-dependent thermal dynamics. 
In Figs.\,\ref{fig:mscr1}B--F, the corrected PS masses remain below the directly comparable LQCD benchmarks.
With the present PS baseline, this discrepancy limits the quantitative accuracy of the extension and motivates further assessment of channel-dependent effects in the CI baseline and the medium response.

The CI curves reveal one further baseline feature. In the light/strange and $s\bar c$ sectors, the SC and AX screening masses turn upward at nearby values of $x$, most clearly in the AX channel. Within the present baseline, this suggests that in heavy-light systems the lighter valence component largely controls the onset of thermal screening. At the same time, the heavier constituent delays the final approach to parity-partner convergence and quasi-free screening. These observations motivate using the $T=0$ reduced dressed-quark mass $M_R=2M_fM_g/(M_f+M_g)$ as the interpolation variable for estimating $x^\ast_{fg}$ in sectors not yet accessible with comparable LQCD precision.



\section{Predictions for future LQCD benchmarks}
\label{sec3}

\begin{figure*}[t]
\centering
\setlength{\panelwidth}{0.32\linewidth}
\setlength{\panelgap}{%
  \dimexpr(\linewidth-3\panelwidth)/2\relax
}
\makebox[\linewidth][c]{%
  \panelplot{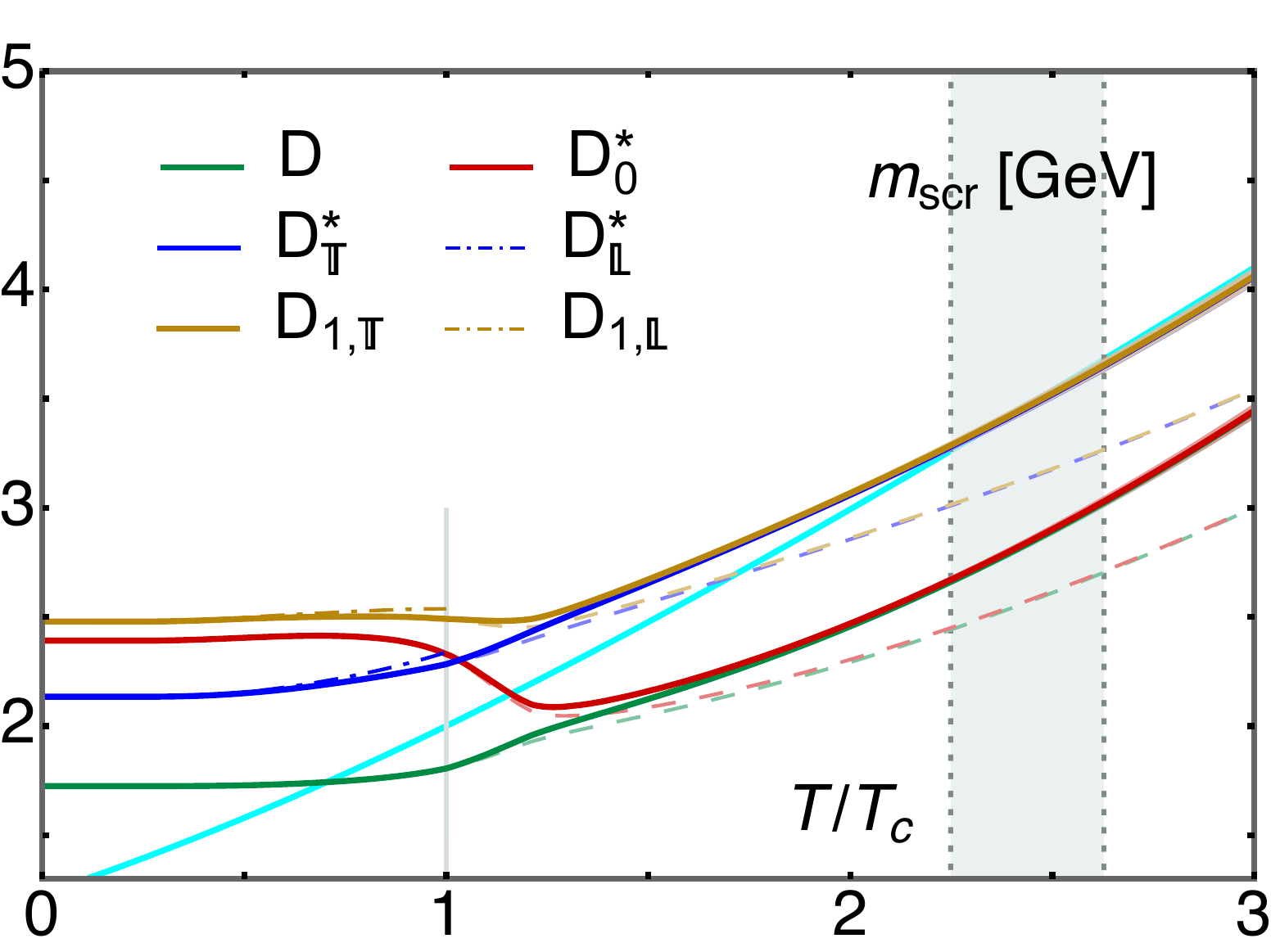}{A}%
  \hspace{\panelgap}%
  \panelplot{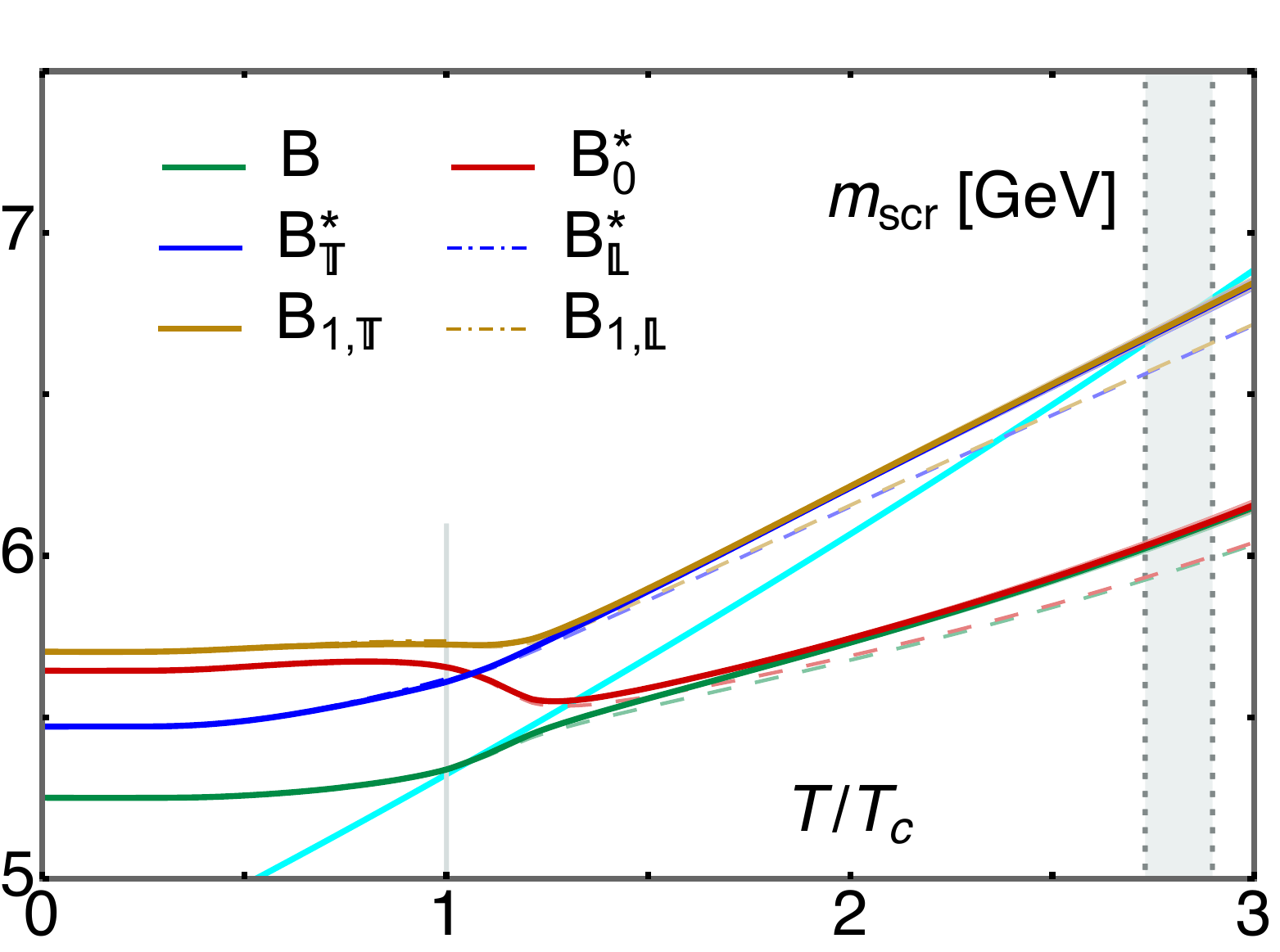}{B}%
  \hspace{\panelgap}%
  \panelplot{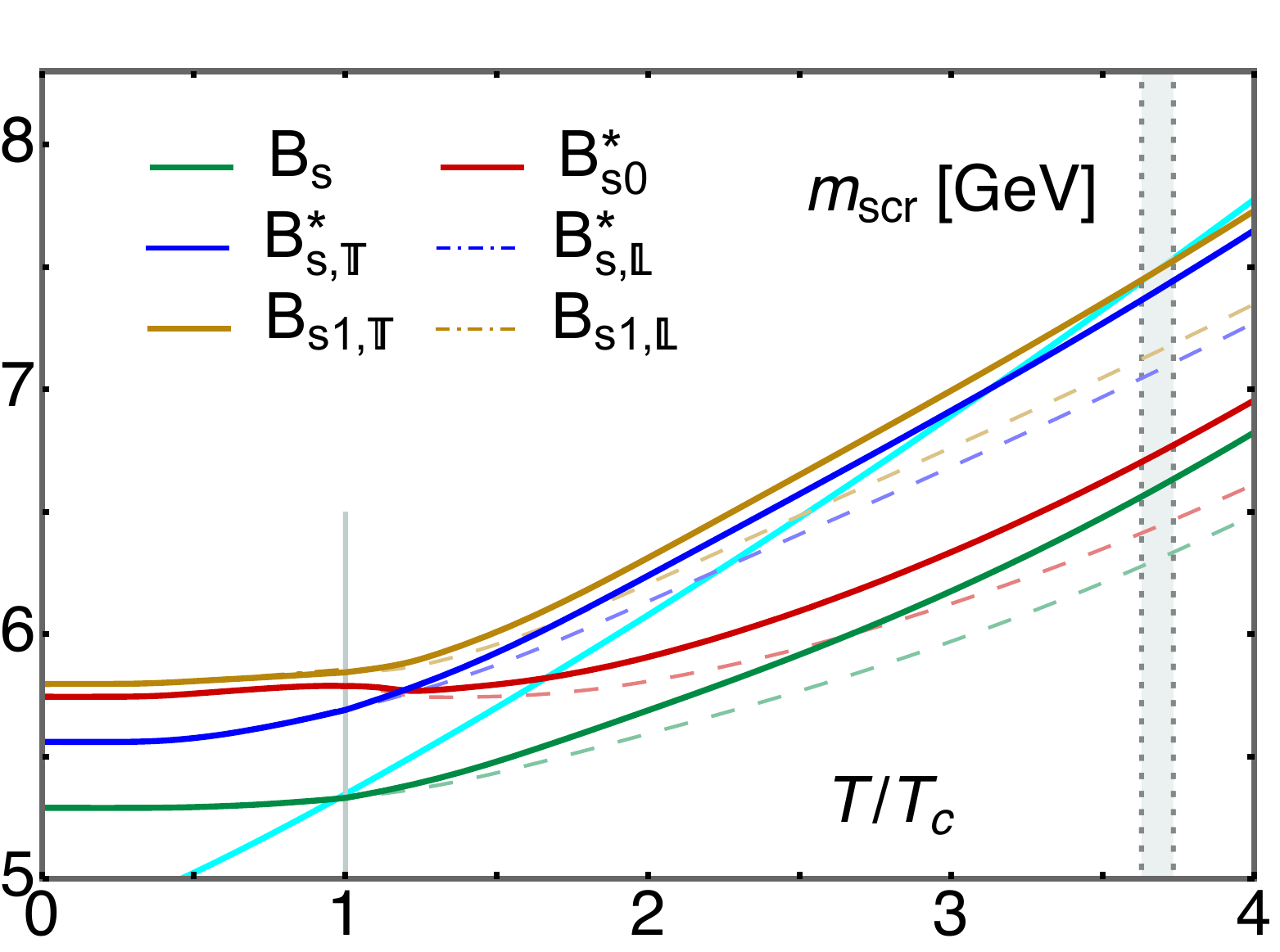}{C}%
}
\par\vspace{0.4em}
\makebox[\linewidth][c]{%
  \panelplot{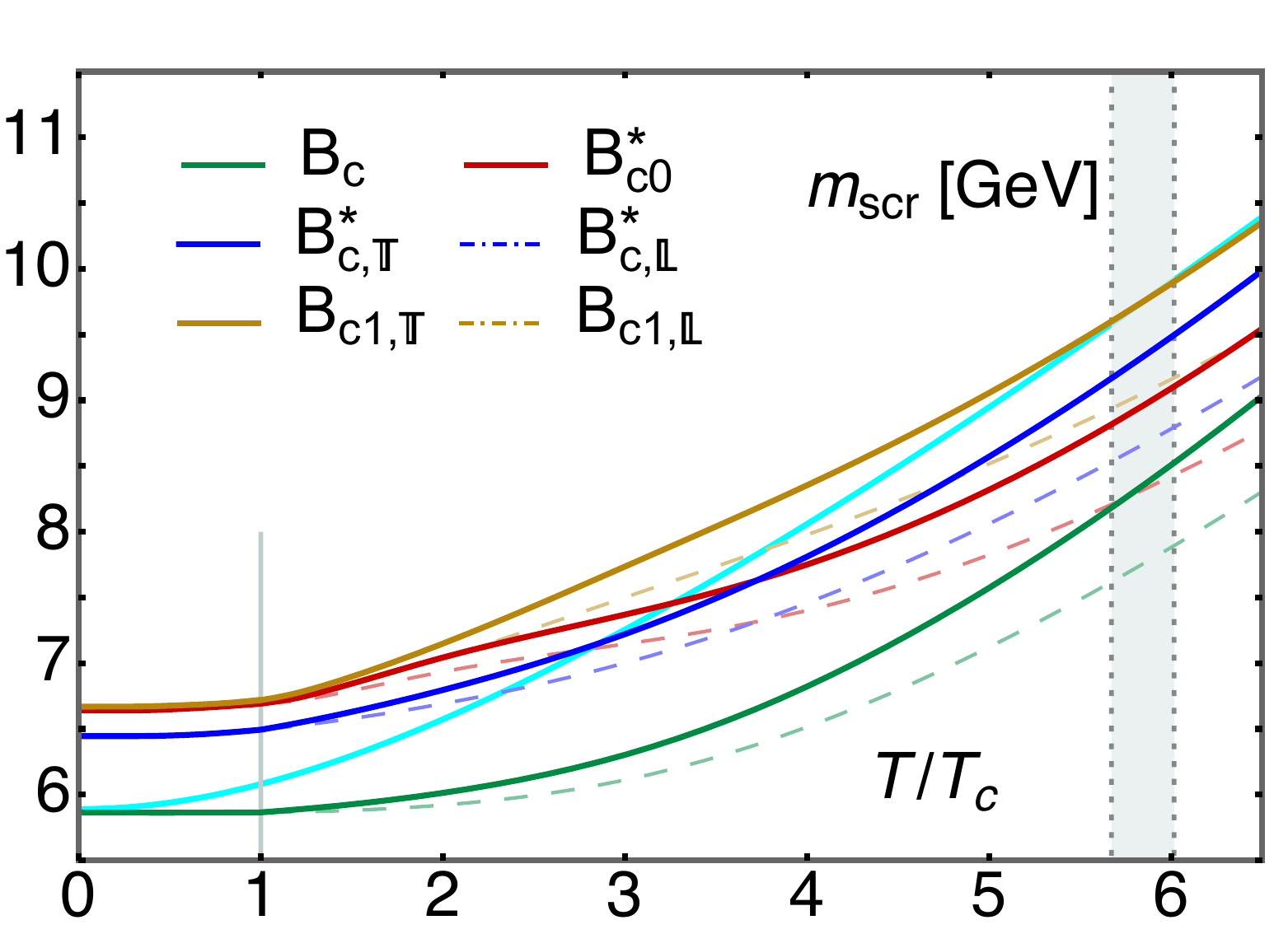}{D}%
  \hspace{\panelgap}%
  \panelplot{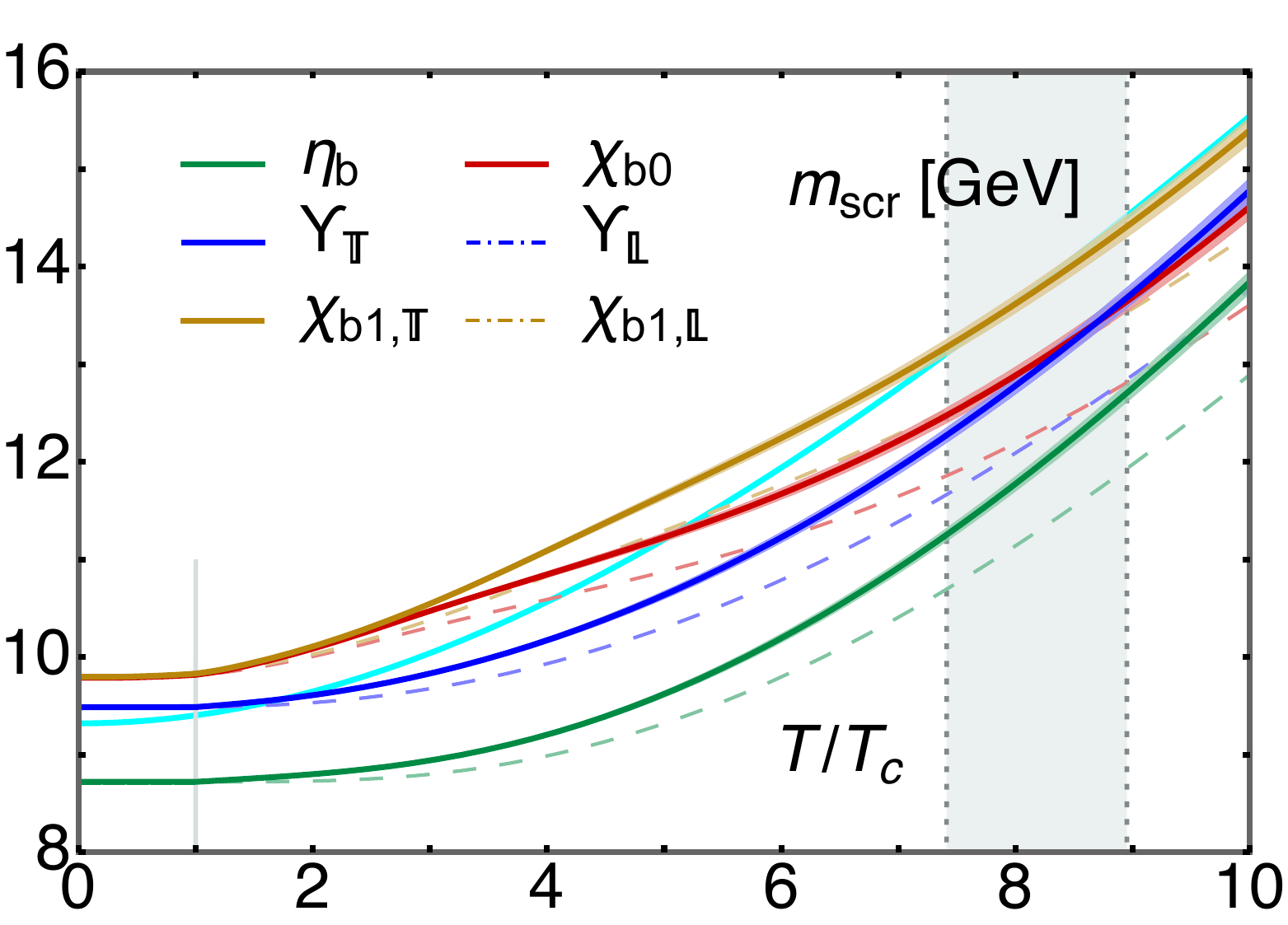}{E}%
}
\caption{\label{fig:mscr2}
Predicted screening spectra in flavor sectors without precision LQCD benchmarks. {\sf Panels A}--{\sf D} show $u\bar c$, $u\bar b$, $s\bar b$, and $c\bar b$ sectors; {\sf panel E} shows $b\bar b$, whose reconstructed quasi-free onset constrains the large-$M_R$ end of the interpolation. Curve identities are indicated in the panels. For $J=1$ states, $\mathds T$ and $\mathds L$ denote transverse and longitudinal modes; longitudinal branches are dot-dashed and have stable CI solutions only for $T<T_c$. 
In all panels, lighter dashed curves are uncorrected CI baselines, while the solid curves are the corrected spectra, obtained from Eq.\,\eqref{corrmass}. 
Gray vertical bands indicate the uncertainties in the AX quasi-free onsets listed in Table\,\ref{tablebvspm}, obtained from the SPM interpolation in {\sf panels A}--{\sf D} and from the bottomonium reconstruction in {\sf panel E}.
Colored bands around the corrected curves propagate this uncertainty to $\beta_{fg}$ and to the screening spectra. 
The PS--SC curves represent an exploratory extension of the AX-calibrated response.
Cyan curves are the CI free-field limits.
}
\end{figure*}

We now predict screening spectra in flavor sectors without precision LQCD benchmarks: $u\bar c$, $u\bar b$, $s\bar b$, and $c\bar b$. The $b\bar b$ sector is treated separately because it provides the only currently available bottomonium information that constrains the large-$M_R$ end of the interpolation. 
As detailed in Sec.\,\ref{supsec5} of the Supplemental Material, we reconstruct a high-temperature AX quasi-free onset from the screening ratios and splittings of Ref.\,\cite{Petreczky:2021zmz}, using its $h_b(1^{+-})$ information as a phenomenological proxy for the $\chi_{b1}(1^{++})$ channel. This reconstructed onset is used as a systematic input, not as a direct precision AX benchmark.

We estimate $x^\ast_{fg}(M_R)$ using a Monte-Carlo SPM reconstruction\,\cite{Schlessinger:1966zz,Schlessinger:1968vsk,Tripolt:2016cya,Chen:2018nsg,Cui:2020rmu,Cui:2021vgm}; see Sec.\,\ref{supsec6} of the Supplemental Material for details. The light/strange-sector quasi-free onset is sampled over $M_R^\ell\in[M_u,M_s]$, while the reconstructed bottomonium onset is sampled over $x^\ast_{b\bar b}=8.18\pm0.77$. The $s\bar c$ and $c\bar c$ onsets are kept at their central values. Only pole-free, monotonic interpolants are retained. The envelopes obtained from $10^3$, $5\times10^3$, and $10^4$ samples are stable; the final band is the envelope of the accepted $10^4$-sample ensemble. 
This band is a systematic interpolation uncertainty, not a statistical lattice error. 
With the $b\bar b$ onset included, the light-charm and open-bottom sectors are constrained by interpolation in $M_R$, rather than by a long extrapolation beyond charm.

\begin{table}[t]
\caption{\label{tablebvspm}
Predicted quasi-free onsets and correction parameters. The values of $x^\ast_{fg}$ are obtained from the Monte-Carlo SPM interpolation in the reduced mass $M_R$, except for the underlined $b\bar b$ entry, which is the reconstructed bottomonium quasi-free onset used as input. 
The corresponding $M_R$ values are listed in Sec.\,\ref{supsec6} of the Supplemental Material. 
For each sector, $\beta_{fg}$ is fixed by Eq.\,\eqref{xast2}.
}
\begin{ruledtabular}
\begin{tabular}{lccccc}
Flavor sector
& $u\bar c$ & $u\bar b$ & $s\bar b$ & $c\bar b$ & $b\bar b$ \\
\hline
$x^\ast_{fg}$ & $2.46(18)$ & $2.83(8)$ & $3.66(5)$ & $5.78(17)$ & $\underline{8.18(77)}$ \\
$\beta_{fg}$ & $0.14(1)$ & $0.019(3)$ & $0.034(1)$ & $0.031(1)$ & $0.016(2)$ \\
\end{tabular}
\end{ruledtabular}
\end{table}

The predicted $x^\ast_{fg}$ values and corresponding $\beta_{fg}$ are listed in Table\,\ref{tablebvspm}. 
For each accepted SPM interpolant, the corresponding $\beta_{fg}$ is fixed by
\begin{align}
\label{xast2}
m_{\mathrm{scr},\mathrm{AX}}^{\mathrm{corr},fg}(x^\ast_{fg})
=
m_{\rm free}^{\mathrm{CI},fg}(x^\ast_{fg})\,,
\end{align}
and then applied without readjustment to PS, SC, VC, and AX. Eq.\,\eqref{xast2} is the prediction-sector analogue of the AX onset matching used in the benchmark sectors. 

Fig.\,\ref{fig:mscr2} shows the predicted VC--AX spectra and the exploratory PS--SC extension on $1\le x\lesssim x^\ast_{fg}$.
The heavy-light sectors $u\bar c$, $u\bar b$, and $s\bar b$ share a common pattern: SC and AX screening masses turn upward at nearby values of $x$, most clearly in AX, around $x\simeq1.25$. This scale is close to that observed in the light/strange sectors and in $s\bar c$, supporting the interpretation that the lighter valence component controls the onset of thermal screening.

By contrast, $c\bar b$ and $b\bar b$ resemble $c\bar c$: heavier valence constituents delay parity-partner convergence and quasi-free screening because current-quark masses explicitly break chiral symmetry. Near $T_c$, the AX and SC branches no longer show the typical decrease-then-increase pattern seen in light/strange and heavy-light systems, reflecting the limitation of the simple spin--orbit ansatz, Eqs.\,\eqref{gsom}, for heavy-heavy AX and SC channels. Once this ansatz-sensitive region is passed, the spectra approach the parity-convergent, quasi-free branch governed by $\mathfrak R$. These spectra therefore provide concrete targets for future finite-temperature LQCD studies of light-charm and bottom-containing meson screening.



\section{Summary}
\label{sec4}

We have identified a common finite-interval response $\mathfrak R(x;\beta_{fg})$ that organizes the available light/strange and charm-containing AX lattice benchmarks relative to a symmetry-preserving DSE/CI baseline.
One AX datum per benchmark sector fixes $\beta_{fg}$; the remaining AX data test the temperature dependence of $\mathfrak R$, while the VC benchmarks provide an independent non-input validation.
The matching endpoints are operational scales defined by screening-mass proximity to the corresponding free-field references.

The extension to the ordinary CI PS--SC sector remains exploratory and does not quantitatively reproduce the available PS lattice benchmarks with the present baseline.

Within this baseline, the lighter valence component largely controls the initial upturn of SC and AX screening masses in heavy-light systems, whereas heavier constituents delay parity-partner convergence and quasi-free screening.
An SPM interpolation in the reduced dressed-quark mass, constrained by the benchmark onsets and a reconstructed bottomonium onset included as a systematic input, yields lattice-testable onset scales and VC--AX screening spectra for light--charm and open-bottom mesons.



\section*{Acknowledgments}
We are grateful for constructive communications with Heng-Tong Ding, Christian S.~Fischer, Craig D.~Roberts, and Qun Wang.
Work supported by: National Natural Science Foundation of China (Grants No. 12247103, No. 12305134).
%


\medskip

\newpage


\clearpage

\section*{Supplemental Material}

\setcounter{secnumdepth}{1}

\setcounter{section}{0}
\setcounter{equation}{0}
\setcounter{figure}{0}
\setcounter{table}{0}

\renewcommand{\thesection}{S\arabic{section}}
\renewcommand{\theequation}{S\arabic{equation}}
\renewcommand{\thefigure}{S\arabic{figure}}
\renewcommand{\thetable}{S\arabic{table}}

\providecommand{\theHsection}{}
\renewcommand{\theHsection}{supp.sec.\arabic{section}}
\renewcommand{\theHequation}{supp.eq.\arabic{equation}}
\renewcommand{\theHfigure}{supp.fig.\arabic{figure}}
\renewcommand{\theHtable}{supp.tab.\arabic{table}}


\section{Dyson--Schwinger framework}
\label{supsec1}

At finite temperature $T$, the Bethe--Salpeter amplitude (BSA) $\Gamma^{J^P}_{[f\bar g]}(p_{\omega_p},P_0)$ for a meson composed of a valence quark of flavor $f$ and a valence antiquark of flavor $\bar g$ is obtained from the homogeneous Bethe--Salpeter equation (BSE)
\begin{align}
\label{eqbse}
\big[\Gamma^{J^P}_{[f\bar g]}(p_{\omega_p},P_0)\big]_{tu}
= {} & T\sum_{n_q=-\infty}^{\infty}
\int \frac{d^3\vec q}{(2\pi)^3}\,
\big[\chi^{J^P}_{[f\bar g]}(q_{\omega_q},P_0)\big]_{sr}
\nonumber\\
&\times
\mathscr K^{tu}_{rs}(p_{\omega_p},q_{\omega_q},P_0)\,.
\end{align}
Here $J^P$ labels the quantum-number channel, $p_{\omega_p}=(\vec p,\omega_p)$, with $\omega_p=(2n_p+1)\pi T$ the fermion Matsubara frequency, and $P_0=(\vec P,0)$ is the total meson momentum with vanishing Matsubara frequency. The indices $r,\ldots,u$ collectively denote color, Dirac, and flavor degrees of freedom. The Bethe--Salpeter wave function is
\begin{align}
\chi^{J^P}_{[f\bar g]}(q_{\omega_q},P_0)
=
S_f(q^+_{\omega_q})\,
\Gamma^{J^P}_{[f\bar g]}(q_{\omega_q},P_0)\,
S_{\bar g}(q^-_{\omega_q})\,,
\end{align}
where $S_f$ is the dressed-quark propagator of flavor $f$, and $q^\pm_{\omega_q}=q_{\omega_q}\pm P_0/2$. The quantity $\mathscr K^{tu}_{rs}$ denotes the Bethe--Salpeter kernel. Eq.\,\eqref{eqbse} is solved as an eigenvalue problem, and the screening mass $m_{\rm scr}(T)$ is determined from\,\cite{Wang:2013wk,Chen:2024emt,Maris:2000ig,Blaschke:2000gd,Gao:2020hwo}
\begin{align}
P_0^2=\vec P^{\,2}=-m_{\rm scr}^2(T)\,.
\end{align}

The leading-order term in the symmetry-preserving truncation scheme of Refs.\,\cite{Munczek:1994zz,Bender:1996bb} is the rainbow--ladder (RL) truncation. For the BSE in Eq.\,\eqref{eqbse}, this amounts to
\begin{align}
\label{rlkernel}
\mathscr K^{tu}_{rs}
=
\hat{\mathscr T}_{\mu\nu}
\bigg[\frac{\lambda^a}{2}i\gamma_\mu\bigg]_{ts}
\bigg[\frac{\lambda^a}{2}i\gamma_\nu\bigg]_{ru}\,,
\end{align}
where $\hat{\mathscr T}_{\mu\nu}$ encodes the dressed-gluon propagator and the effective interaction strength. With a suitable effective interaction, RL provides a reliable description of channels in which corrections beyond leading order largely cancel, in particular flavor-nonsinglet pseudoscalar (PS) and vector (VC) ground states\,\cite{Chang:2009zb,Fischer:2009jm,Williams:2015cvx,Qin:2020jig}. The RL kernel preserves the relevant Ward--Takahashi identities, but it does not contain the anomalous contribution required for anomaly-dominated channels\,\cite{Maris:2000ig,Bhagwat:2007ha,Ding:2018xwy}. Hence, throughout this work, RL is used to study ordinary chiral restoration, or, away from the chiral limit, parity-partner convergence, rather than effective $U_A(1)$ restoration.

The axial-vector Ward--Takahashi identity constrains the Bethe--Salpeter kernel and the dressed-quark self-energy to be mutually consistent\,\cite{Roberts:1994dr,Roberts:2000aa,Eichmann:2016yit,Fischer:2018sdj}.
In the RL truncation, the dressed-quark propagator is obtained from the gap equation
\begin{subequations}
\label{gapeq}
\begin{align}
S_f^{-1}(p_{\omega_p})
&=
 i\vec\gamma\!\cdot\!\vec p
+i\gamma_4\omega_p
+m_f
+\Sigma_f(p_{\omega_p})\,,
\\
\Sigma_f(p_{\omega_p})
&=
T\sum_{n_q=-\infty}^{\infty}
\int\frac{d^3\vec q}{(2\pi)^3}\,
\hat{\mathscr T}_{\mu\nu}
\frac{\lambda^a}{2}\gamma_\mu
S_f(q_{\omega_q})
\frac{\lambda^a}{2}\gamma_\nu\,,
\end{align}
\end{subequations}
where $m_f$ is the current-quark mass of flavor $f$; the chiral limit is defined by $m_f=0$.


\section{Contact interaction: details}
\label{supsec2}

\begin{table}[t]
\caption{\label{tableqmasses}
Single-flavor parameter sets and the corresponding $T=0$ pseudoscalar masses, $m_{0^-}$, and leptonic decay constants, $f_{0^-}$. All dimensionful quantities are quoted in GeV.
}
\begin{ruledtabular}
\begin{tabular}{lcccccc}
Flavor &
$\alpha_{\mathrm{IR}}/\pi$ &
$\Lambda_{\mathrm{UV}}$ &
$m_f$ &
$M_f$ &
$m_{0^-}$ &
$f_{0^-}$
\\
\hline
chiral limit & 0.36  & 0.91 & 0     & 0.36 & 0    & 0.09 \\
$u,d$      & 0.36  & 0.91 & 0.007 & 0.37 & 0.14 & 0.10 \\
$s$        & 0.36  & 0.91 & 0.17  & 0.53 & 0.50 & 0.11 \\
$c$        & 0.053 & 1.89 & 1.23  & 1.60 & 2.98 & 0.24 \\
$b$        & 0.012 & 3.54 & 4.66  & 4.83 & 9.40 & 0.41
\end{tabular}
\end{ruledtabular}
\end{table}

The contact interaction (CI) used in this work is a Nambu--Jona-Lasinio
(NJL)-type realization of the RL kernel:
\begin{align}
\label{ci}
\hat{\mathscr T}_{\mu\nu}^{\mathrm{CI}}
=
\frac{4\pi\alpha_{\mathrm{IR}}}{m_G^2}\,
\delta_{\mu\nu}\,.
\end{align}
The model is specified by the current-quark mass $m_f$, the gluon mass scale $m_G$, the infrared and ultraviolet cutoffs $\Lambda_{\mathrm{IR}}$ and $\Lambda_{\mathrm{UV}}$, and the effective coupling $\alpha_{\mathrm{IR}}$. We set $m_G=0.5\,\mathrm{GeV}$, consistent with QCD studies of the gluon mass scale\,\cite{Binosi:2016nme}. The infrared cutoff is fixed at $\Lambda_{\mathrm{IR}}=0.24\,\mathrm{GeV}\sim\Lambda_{\mathrm{QCD}}$; in the proper-time regularization, this implements confinement by removing quark production thresholds\,\cite{Ebert:1996vx,Krein:1990sf}. The ultraviolet cutoff $\Lambda_{\mathrm{UV}}$ sets the momentum domain over which the CI is taken to be applicable, while $\alpha_{\mathrm{IR}}$ fixes the overall interaction strength.

The single-flavor parameter sets in Table\,\ref{tableqmasses} are fixed by $T=0$ PS observables\,\cite{Yin:2021uom}. For mixed-flavor mesons, the corresponding values of $\Lambda_{\mathrm{UV}}$ and $\alpha_{\mathrm{IR}}$ are obtained using the same prescription as in Ref.\,\cite{Yin:2021uom}. This prescription implements the physical constraint that hadrons with larger momentum-space support are described by a weaker effective coupling.

\begin{figure}[t]
\centering
\includegraphics[clip,width=0.4\textwidth]{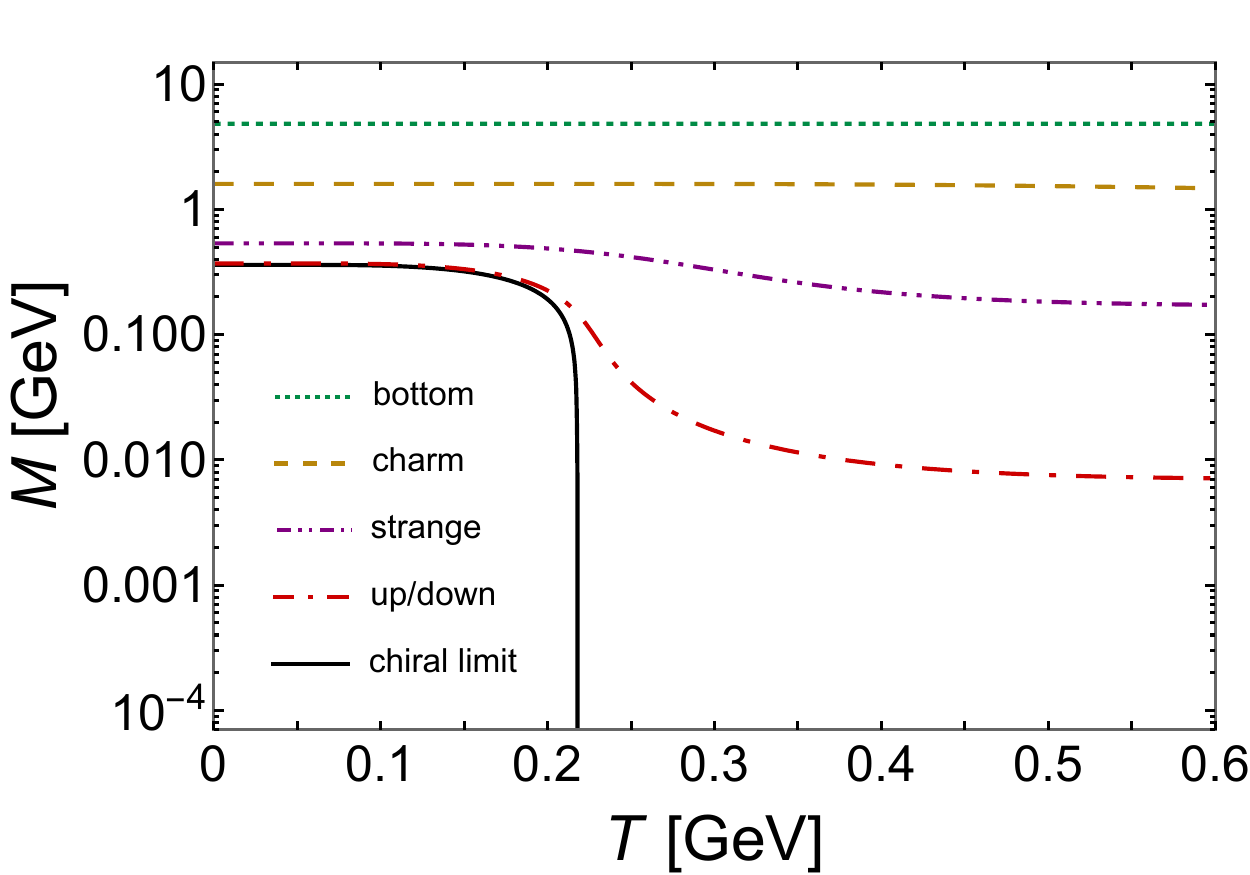}
\caption{
Temperature dependence of the dressed-quark mass $M_f(T)$ in the chiral limit and for physical current-quark masses. The vertical axis is logarithmic. Thermal effects suppress DCSB at high temperature, and the thermal response weakens with increasing current-quark mass.
\label{figMT}
}
\end{figure}

The gap equation determines the thermal dressed-quark masses that enter
the BSEs. With the CI kernel in Eq.\,\eqref{ci},
Eq.\,\eqref{gapeq} yields
\begin{align}
\label{cigap}
S_f^{-1}(p_{\omega_p})
=
 i\vec\gamma\!\cdot\!\vec p
+i\gamma_4\omega_p
+M_f(T)\,,
\end{align}
where the contact nature of the interaction makes the dressed-quark mass $M_f(T)$ momentum independent. Using the parameter sets in Table\,\ref{tableqmasses}, we solve the finite-temperature gap equation; the resulting $M_f(T)$ are shown in Fig.\,\ref{figMT}.

In the chiral limit, chiral symmetry is restored through a second-order transition at $T_c^0=0.215\,\mathrm{GeV}$. For physical $u/d$ quarks, this transition becomes a crossover. We identify the pseudocritical temperature, $T_c=0.197\,\mathrm{GeV}$, from the coalescence of the Nambu-negative and Wigner branches\,\cite{Wang:2013wk,Chen:2024emt}. This value is higher than that obtained in modern lattice QCD (LQCD) simulations\,\cite{Bazavov:2011nk}, as expected for NJL-type contact models with temperature-independent couplings, which typically yield $T_c\simeq0.2\,\mathrm{GeV}$ unless additional temperature-dependent parameters are introduced\,\cite{Fukushima:2013rx,Buballa:2003qv}. 
Comparisons with LQCD are therefore made at the same reduced temperature $x=T/T_c$, using the normalized screening masses defined in the main text.

Fig.\,\ref{figMT} also shows that the temperature dependence of $M_f(T)$ weakens rapidly as the current-quark mass increases. At high temperature, thermal effects suppress dynamical chiral symmetry breaking (DCSB), and $M_f(T)$ approaches the corresponding current-quark mass. This trend is consistent with Table\,\ref{tableqmasses}, which shows that the relative dynamical contribution to the dressed-quark mass decreases from light to heavy flavors.


\begin{table*}[t]
\caption{\label{tab:mdsmasses}
Computed $T=0$ masses, in GeV, for ground-state $J^P=0^\mp,1^\mp$ mesons in the present CI setup. The underlined entries are the PS masses used, together with the corresponding decay constants in Table\,\ref{tableqmasses}, to fix the interaction strengths, current-quark masses, and ultraviolet cutoffs. Mixed-flavor parameters are obtained using the prescription of Ref.\,\cite{Yin:2021uom}. The CI entries listed here are recomputed for the present finite-temperature analysis. For PS mesons, the row labeled $F^{0^-}\to0$ gives the result obtained after removing the pseudovector covariant in Eq.\,\eqref{bsamps}. Reference masses are taken from the PDG\,\cite{ParticleDataGroup:2024cfk} where available; $m_{B_c^\ast}$, $m_{B_{c0}^\ast}$, and $m_{B_{c1}}$ are from Ref.\,\cite{Mathur:2018epb}, and $m_{\eta_s}$ from Ref.\,\cite{Dowdall:2013rya}. A dash indicates that no empirical or LQCD reference value is available.
}
\begin{ruledtabular}
\begin{tabular*}{\textwidth}{@{\extracolsep{\fill}}lcccccccccc@{}}
$J^P=0^-$ &
$\pi\,(u\bar d)$ &
$K\,(u\bar s)$ &
$\eta_s\,(s\bar s)$ &
$D\,(u\bar c)$ &
$D_s\,(s\bar c)$ &
$\eta_c\,(c\bar c)$ &
$B\,(u\bar b)$ &
$B_s\,(s\bar b)$ &
$B_c\,(c\bar b)$ &
$\eta_b\,(b\bar b)$
\\
\hline
CI, this work $(F^{0^-}\to0)$ &
0.12 & 0.42 & 0.58 & 1.73 & 1.78 & 2.64 & 5.25 & 5.29 & 5.86 & 8.72
\\
CI, this work &
$\underline{0.14}$ & $\underline{0.50}$ & 0.70 & 1.92 & 2.00 &
$\underline{2.98}$ & 5.42 & 5.50 & 6.33 & $\underline{9.40}$
\\
expt./LQCD &
0.14 & 0.50 & 0.69 & 1.87 & 1.97 & 2.98 & 5.28 & 5.37 & 6.28 & 9.40
\\
\hline
\multicolumn{11}{c}{}\\[-0.5em]
\hline
$J^P=1^-$ &
$\rho\,(u\bar d)$ &
$K^\ast\,(u\bar s)$ &
$\phi\,(s\bar s)$ &
$D^\ast\,(u\bar c)$ &
$D_s^\ast\,(s\bar c)$ &
$J/\psi\,(c\bar c)$ &
$B^\ast\,(u\bar b)$ &
$B_s^\ast\,(s\bar b)$ &
$B_c^\ast\,(c\bar b)$ &
$\Upsilon\,(b\bar b)$
\\
\hline
CI, this work &
0.93 & 1.03 & 1.13 & 2.13 & 2.22 & 3.19 & 5.47 & 5.56 & 6.44 & 9.48
\\
expt./LQCD &
0.78 & 0.89 & 1.02 & 2.01 & 2.11 & 3.10 & 5.33 & 5.42 & 6.33 & 9.46
\\
\hline
\multicolumn{11}{c}{}\\[-0.5em]
\hline
$J^P=0^+$ &
$\sigma\,(u\bar d)$ &
$\kappa\,(u\bar s)$ &
$\sigma_s\,(s\bar s)$ &
$D_0^\ast\,(u\bar c)$ &
$D_{s0}^\ast\,(s\bar c)$ &
$\chi_{c0}\,(c\bar c)$ &
$B_0^\ast\,(u\bar b)$ &
$B_{s0}^\ast\,(s\bar b)$ &
$B_{c0}^\ast\,(c\bar b)$ &
$\chi_{b0}\,(b\bar b)$
\\
\hline
CI, this work &
1.22 & 1.34 & 1.46 & 2.39 & 2.50 & 3.46 & 5.64 & 5.74 & 6.64 & 9.79
\\
expt./LQCD &
-- & -- & -- & 2.30 & 2.32 & 3.42 & -- & -- & 6.71 & 9.86
\\
\hline
\multicolumn{11}{c}{}\\[-0.5em]
\hline
$J^P=1^+$ &
$a_1\,(u\bar d)$ &
$K_1\,(u\bar s)$ &
$f_1\,(s\bar s)$ &
$D_1\,(u\bar c)$ &
$D_{s1}\,(s\bar c)$ &
$\chi_{c1}\,(c\bar c)$ &
$B_1\,(u\bar b)$ &
$B_{s1}\,(s\bar b)$ &
$B_{c1}\,(c\bar b)$ &
$\chi_{b1}\,(b\bar b)$
\\
\hline
CI, this work &
1.37 & 1.48 & 1.59 & 2.48 & 2.58 & 3.51 & 5.70 & 5.80 & 6.67 & 9.80
\\
expt./LQCD &
1.23 & 1.25 & 1.43 & 2.42 & 2.46 & 3.51 & 5.73 & 5.83 & 6.74 & 9.89
\end{tabular*}
\end{ruledtabular}
\end{table*}

\begin{figure*}[t]
\centering
\includegraphics[width=0.95\textwidth]{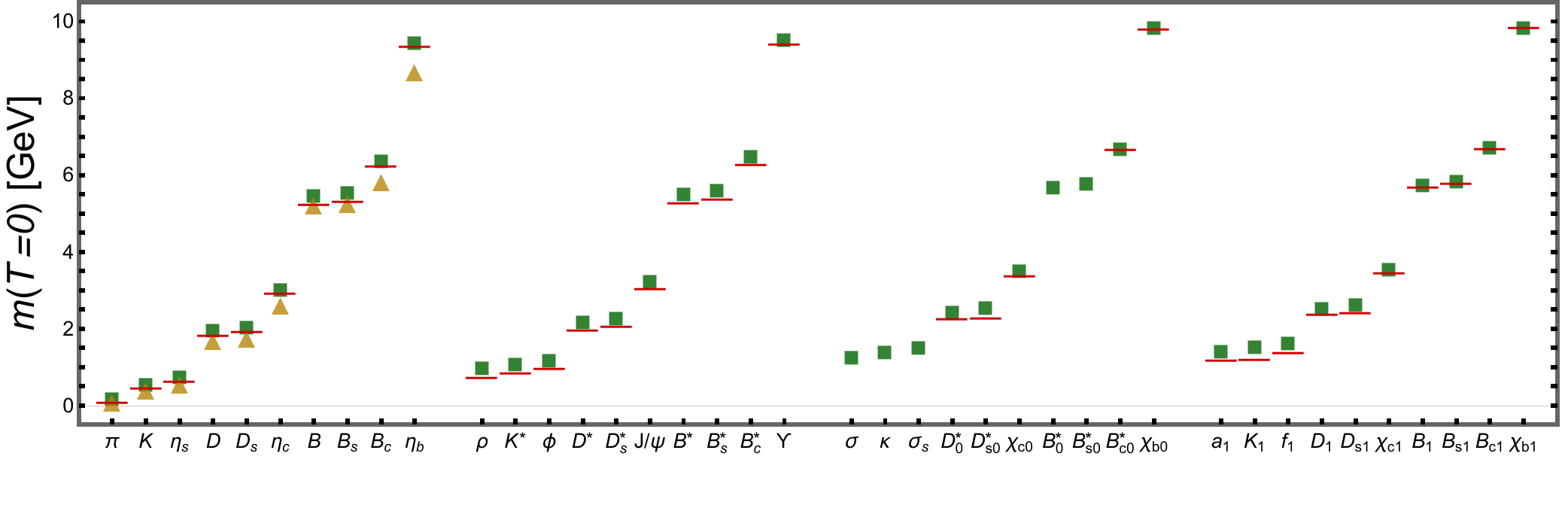}
\caption{
Comparison of $T=0$ meson masses obtained in the present CI calculation with available reference values. Green squares denote the full CI results; red horizontal bars denote experimental values from the PDG\,\cite{ParticleDataGroup:2024cfk}, supplemented by LQCD results for $B_c^\ast$, $B_{c0}^\ast$, and $B_{c1}$\,\cite{Mathur:2018epb}, and for $\eta_s$\,\cite{Dowdall:2013rya}; gold triangles denote PS CI results obtained after setting $F^{0^-}=0$.
\label{fig:massesT0}
}
\end{figure*}


\section{Bethe--Salpeter amplitudes, spin--orbit ansatz, and meson masses at $T=0$}
\label{supsec3}

With the CI kernel in Eq.\,\eqref{ci}, the BSAs are independent of the
relative momentum. For a PS meson, $J^P=0^-$, we use
\begin{align}
\label{bsamps}
\Gamma^{0^-}_{[f\bar g]}(P_0;T)
=
 i\gamma_5 E^{0^-}_{[f\bar g]}(T)
+
\frac{1}{2M_{fg}(T)}
\gamma_5\gamma\!\cdot\!P_0\,
F^{0^-}_{[f\bar g]}(T)\,,
\end{align}
where
\begin{align}
M_{fg}(T)=\frac{M_f(T)M_g(T)}{M_f(T)+M_g(T)}\,.
\end{align}
For its parity partner, the scalar (SC) meson, $J^P=0^+$, the BSA is
\begin{align}
\label{bsamsc}
\Gamma^{0^+}_{[f\bar g]}(P_0;T)
=
\mathbf{1}_D\,E^{0^+}_{[f\bar g]}(T)\,,
\end{align}
where $\mathbf{1}_D$ is the identity matrix in Dirac space.

For $J=1$ channels, the heat bath breaks $O(4)$ symmetry and separates
the longitudinal and transverse components. For VC mesons,
$J^P=1^-$, we write
\begin{subequations}
\label{bsamvc}
\begin{align}
\Gamma^{1^-,\parallel}_{[f\bar g]}(P_0;T)
&=
\gamma_4\,E^{1^-,\parallel}_{[f\bar g]}(T)\,,
\\
\vec\Gamma^{1^-,\perp}_{[f\bar g]}(P_0;T)
&=
\vec\gamma_{\perp}\,E^{1^-,\perp}_{[f\bar g]}(T)\,,
\end{align}
\end{subequations}
with
\begin{align}
(\gamma_\perp)_i=
\left(\delta_{ij}-\frac{P_iP_j}{|\vec P|^2}\right)\gamma_j\,,
\qquad i,j=1,2,3\,.
\end{align}
The corresponding axial-vector (AX) amplitudes, $J^P=1^+$, are
represented analogously:
\begin{subequations}
\label{bsamax}
\begin{align}
\Gamma^{1^+,\parallel}_{[f\bar g]}(P_0;T)
&=
\gamma_5\gamma_4\,E^{1^+,\parallel}_{[f\bar g]}(T)\,,
\\
\vec\Gamma^{1^+,\perp}_{[f\bar g]}(P_0;T)
&=
\gamma_5\vec\gamma_{\perp}\,E^{1^+,\perp}_{[f\bar g]}(T)\,.
\end{align}
\end{subequations}

RL truncation provides a reliable description of flavor-nonsinglet PS and VC ground states, where corrections beyond leading order largely cancel. In SC and AX channels, however, the bare RL kernel does not generate sufficient spin--orbit (SO) repulsion, leading to an inadequate description of the corresponding mass splittings. At $T=0$, a common phenomenological remedy is to multiply the RL kernel in these channels by a constant factor that mimics the missing SO repulsion; see, e.g., Refs.\,\cite{Chen:2012qr,Lu:2017cln,Yin:2021uom,Yin:2019bxe,Cheng:2022jxe}. For the present screening-mass calculation, we follow Ref.\,\cite{Chen:2024emt} and use temperature-dependent factors multiplying the CI kernel in the SC and AX channels:
\begin{subequations}
\label{gsom}
\begin{align}
\label{gsomsc}
\mathfrak g_{\mathrm{SO}}^{q\bar q,0^+}(T)
&=
1-
\frac{M_u(T)-m_u}{M_u(0)-m_u}
\left[1-(0.32)^2\right],
\\
\label{gsomax}
\mathfrak g_{\mathrm{SO}}^{q\bar q,1^+}(T)
&=
1-
\frac{M_u(T)-m_u}{M_u(0)-m_u}
\left[1-(0.25)^2\right]\,.
\end{align}
\end{subequations}
Thus, at $T=0$,
\begin{subequations}
\label{gso}
\begin{align}
\mathfrak g_{\mathrm{SO}}^{q\bar q,0^+}(0)
&=(0.32)^2\,,
\\
\mathfrak g_{\mathrm{SO}}^{q\bar q,1^+}(0)
&=(0.25)^2\,.
\end{align}
\end{subequations}
These are the values used in Refs.\,\cite{Lu:2017cln,Yin:2021uom} to reproduce the phenomenological $a_1-\rho$ and $\sigma-\rho$ inertial-mass splittings. As $T$ increases and DCSB is suppressed, the factors in Eq.\,\eqref{gsom} approach unity. Hence the additional phenomenological SO repulsion is continuously switched off.

\begin{figure*}[t]
\centering
\setlength{\panelwidth}{0.32\linewidth}
\setlength{\panelgap}{%
  \dimexpr(\linewidth-3\panelwidth)/2\relax
}
\makebox[\linewidth][c]{%
  \panelplot{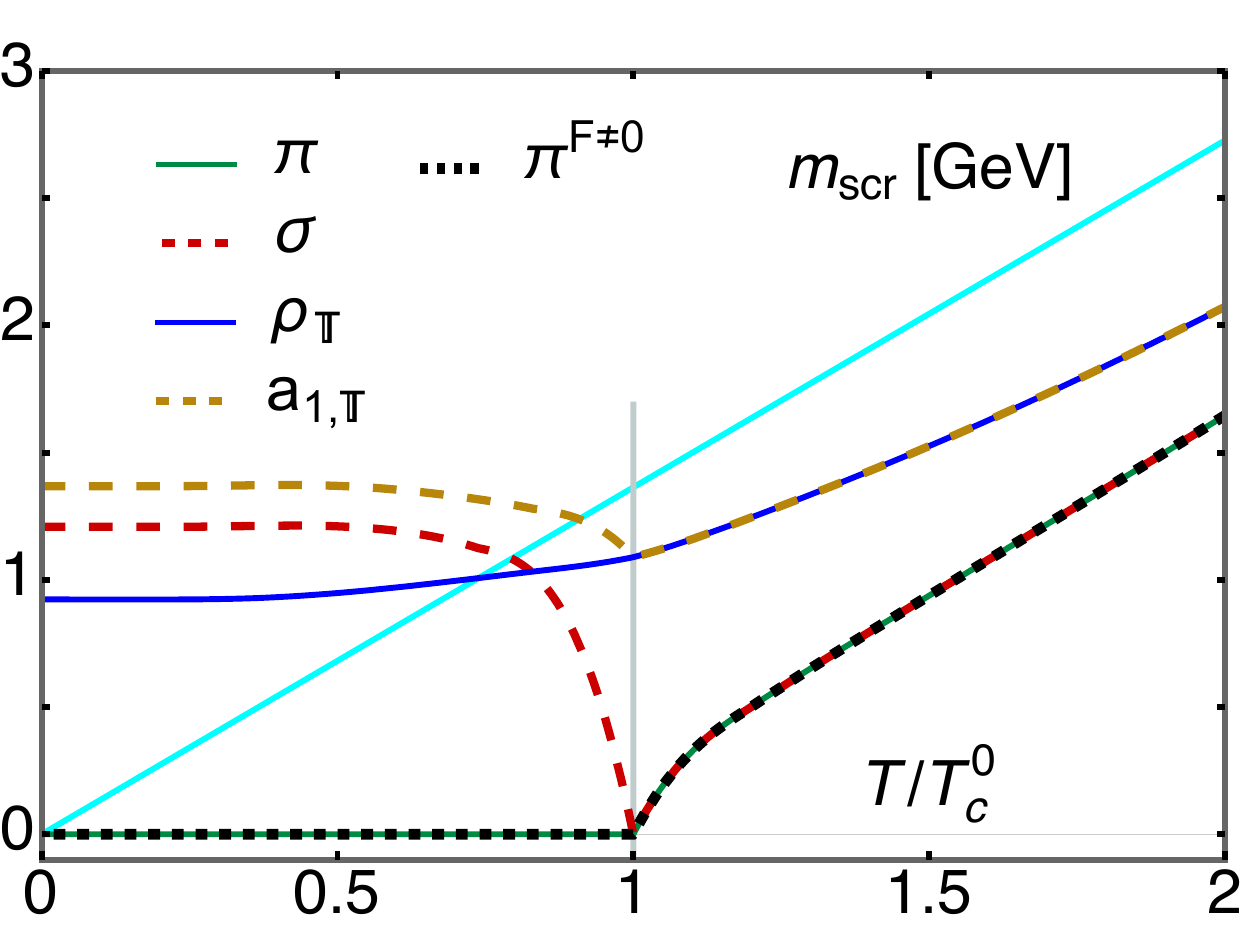}{A}%
  \hspace{\panelgap}%
  \panelplot{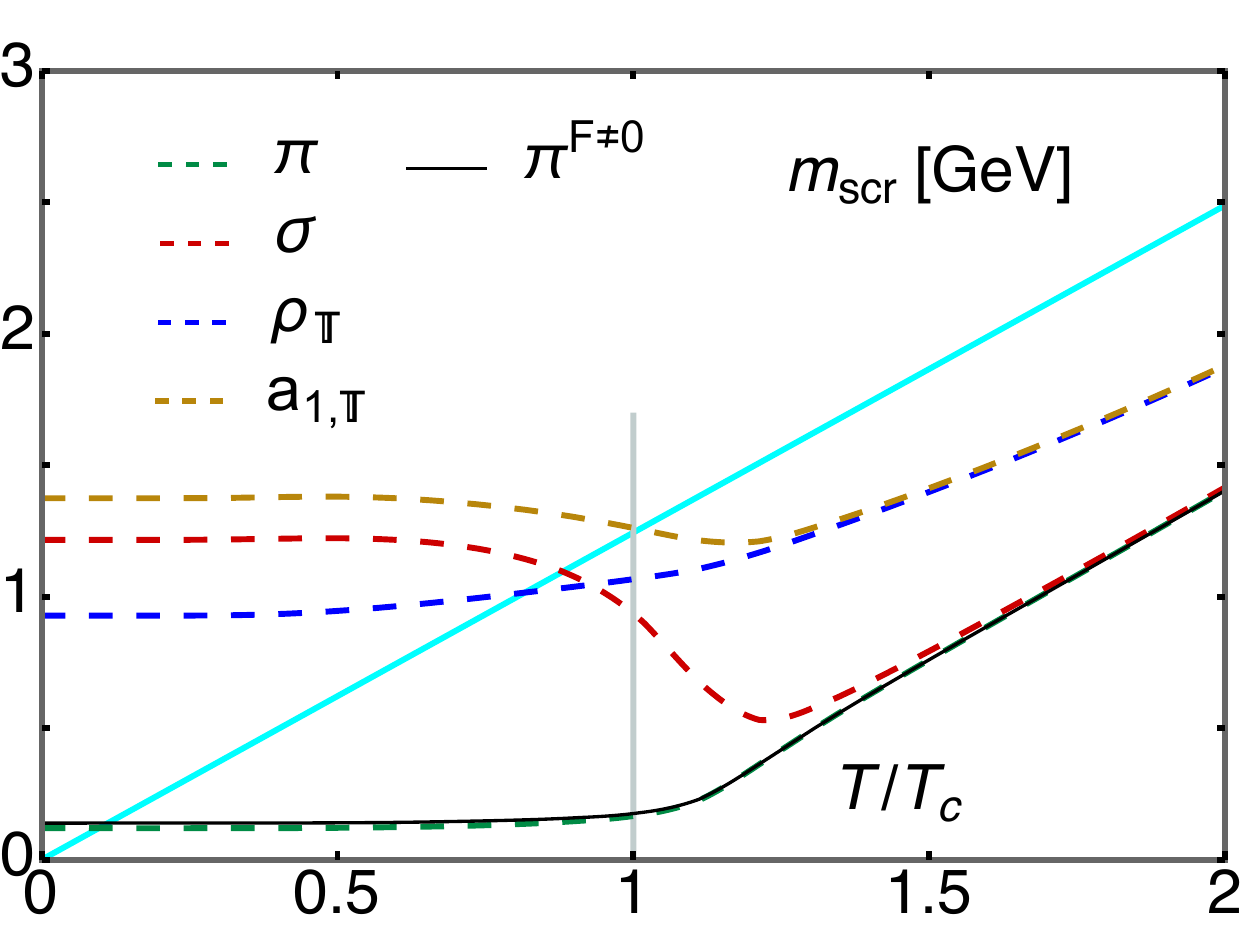}{B}%
  \hspace{\panelgap}%
  \panelplot{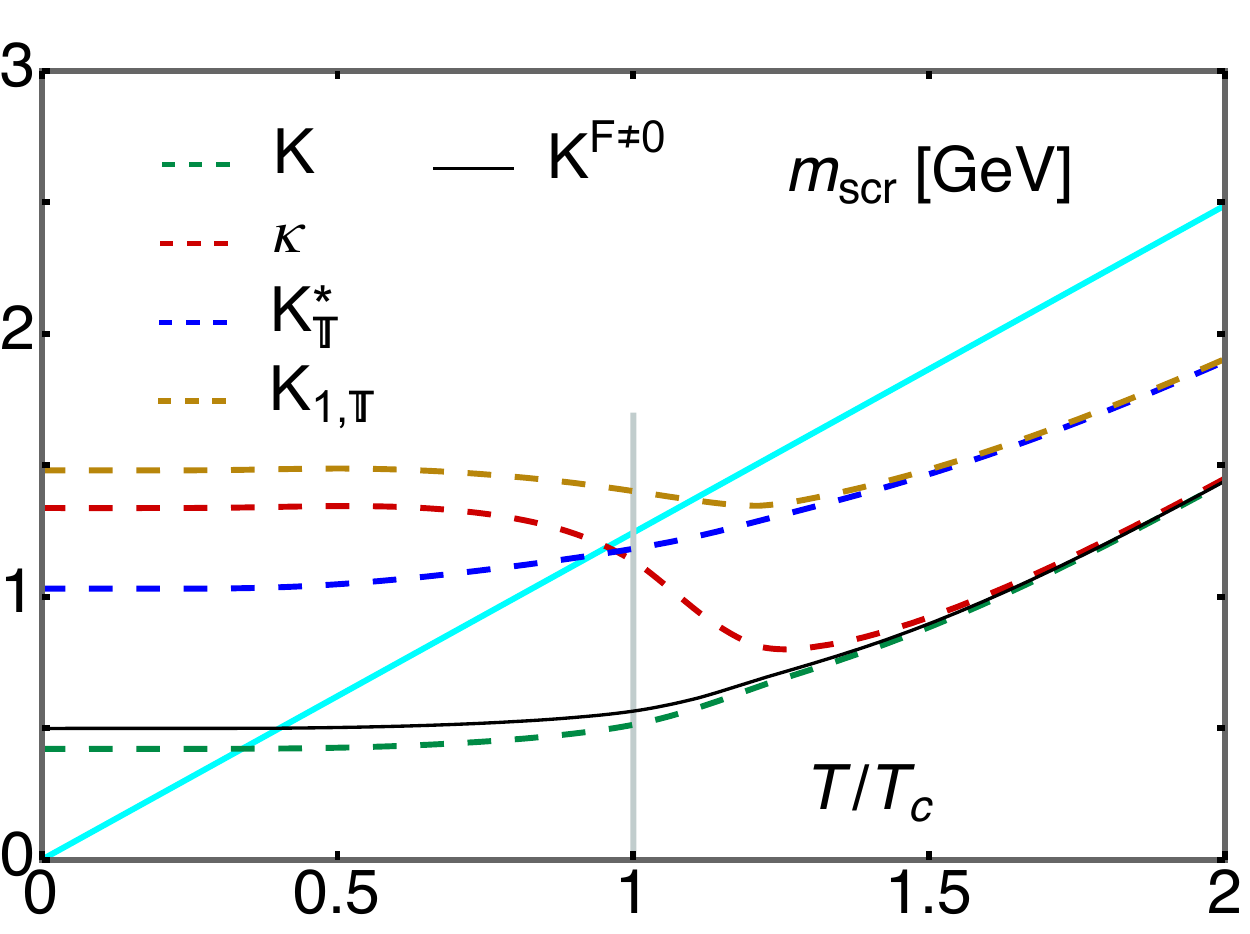}{C}%
}
\par\vspace{0.4em}
\makebox[\linewidth][c]{%
  \panelplot{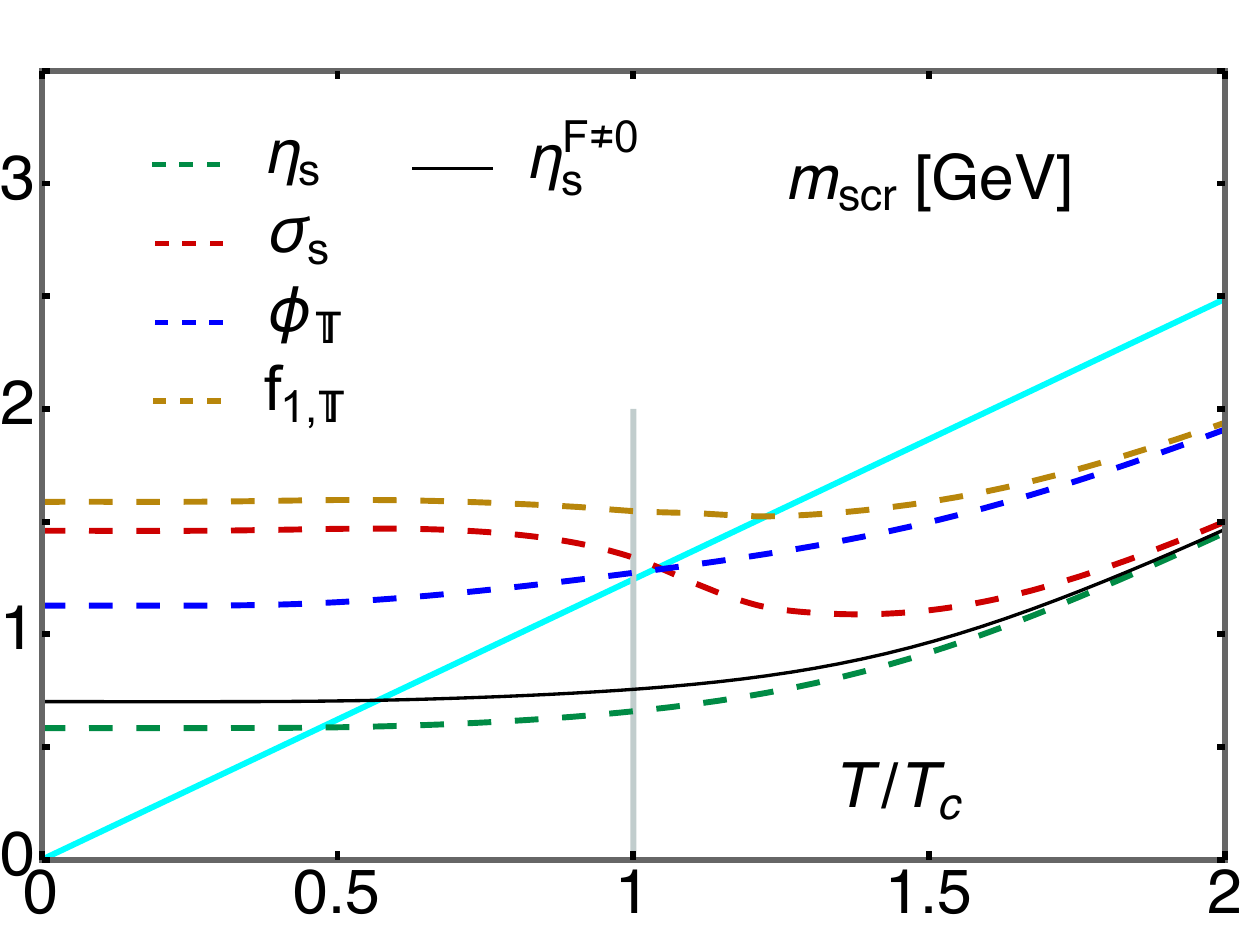}{D}%
  \hspace{\panelgap}%
  \panelplot{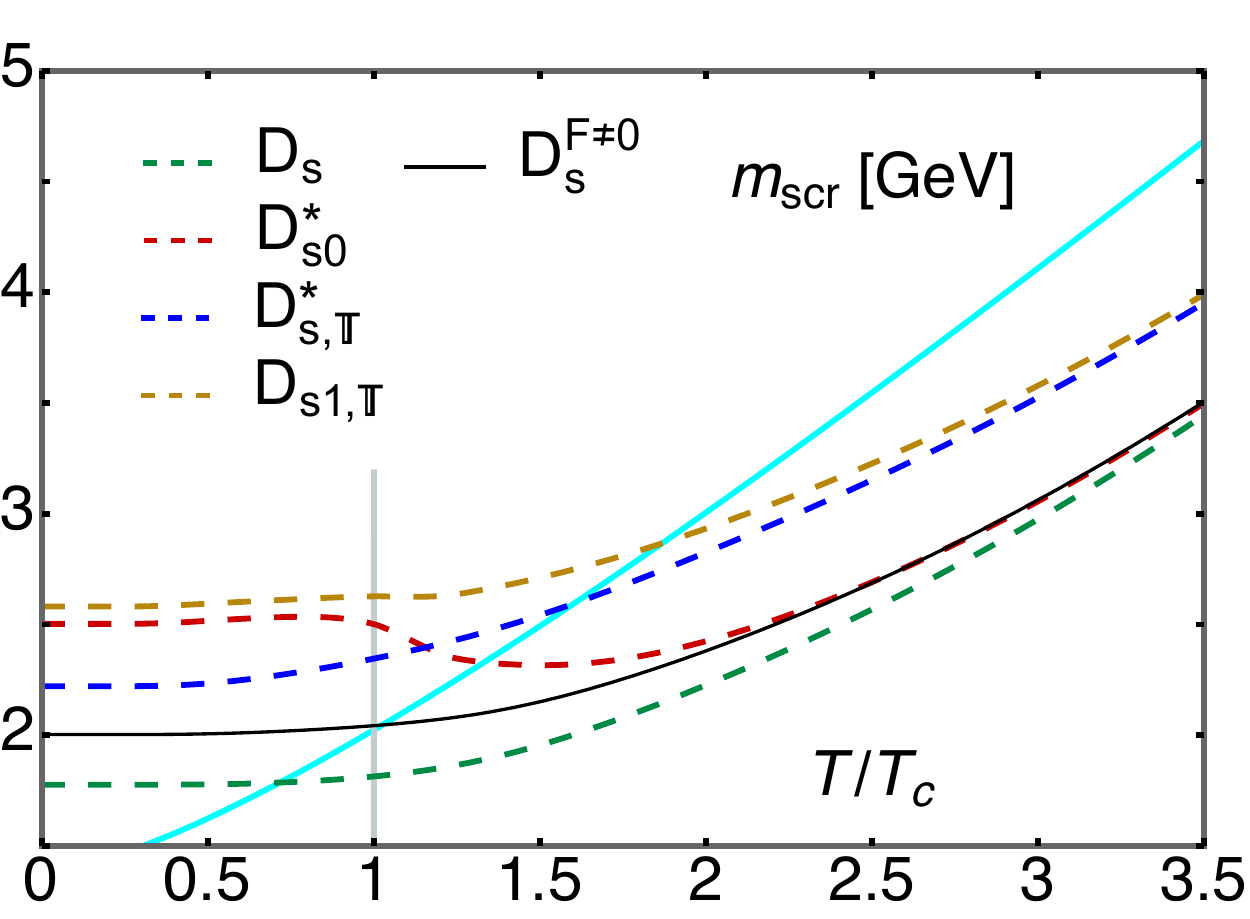}{E}%
  \hspace{\panelgap}%
  \panelplot{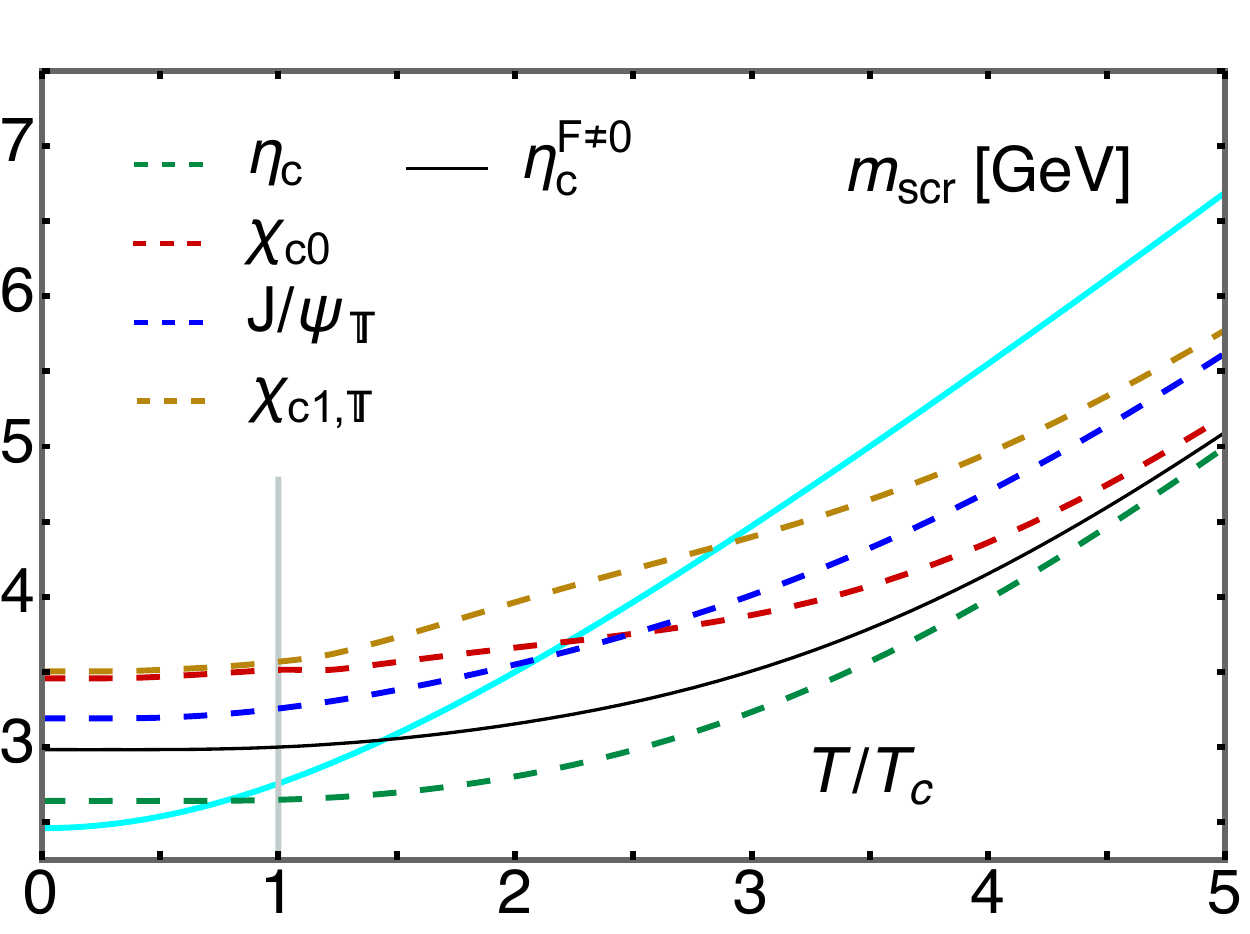}{F}%
}
\caption{\label{fig:supmscr1}
Effect of the PS pseudovector covariant on the uncorrected CI screening spectra in the benchmark flavor sectors. {\sf Panel A} shows the chiral-limit $u\bar d$ sector; {\sf panels B}--{\sf F} show physical $u\bar d$, $u\bar s$, $s\bar s$, $s\bar c$, and $c\bar c$ sectors.
Curve identities are indicated in the panels.
Only transverse $J=1$ modes are shown.
PS curves labeled by the superscript $F\neq0$, e.g., $\pi^{F\neq0}$, are the full solutions with the pseudovector covariant retained; PS curves without this superscript are the reduced $F^{0^-}\to0$ solutions used for the PS--SC comparisons in the main text.
In {\sf panel A}, the full and reduced pion curves coincide within
numerical accuracy.
Cyan curves are the CI free-field limits.
This figure is the uncorrected diagnostic counterpart of Fig.\,\ref{fig:mscr1}.
}
\end{figure*}

With all ingredients fixed, we recompute the $T=0$ spectrum in the present CI setup. We follow the parameter-fixing strategy and mixed-flavor prescription of Ref.\,\cite{Yin:2021uom}, but the masses in Table\,\ref{tab:mdsmasses} are obtained independently for the present finite-temperature study. The values listed here define the $T=0$ baseline used consistently in the screening-mass analysis.

The PS BSA contains a nonzero pseudovector covariant, the $F^{0^-}$-term, Eq.\,\eqref{bsamps}. At $T=0$, this component is quantitatively important: setting $F^{0^-}=0$ lowers all PS masses, as shown in Table\,\ref{tab:mdsmasses} and Fig.\,\ref{fig:massesT0}. 
The analysis of the effect of this term at $T\neq0$ is given in Sec.\,\ref{supsec4}. 

Overall, after excluding the fitted entries, the CI provides a semiquantitative description of the five-flavor ground-state spectrum, with particularly good agreement in the heavy-flavor sector. This establishes the $T=0$ reference point for the finite-temperature screening-mass analysis.


\section{Effect of the pseudovector covariant on finite-temperature PS--SC comparisons}
\label{supsec4}

\begin{figure*}[t]
\centering
\setlength{\panelwidth}{0.32\linewidth}
\setlength{\panelgap}{%
  \dimexpr(\linewidth-3\panelwidth)/2\relax
}
\makebox[\linewidth][c]{%
  \panelplot{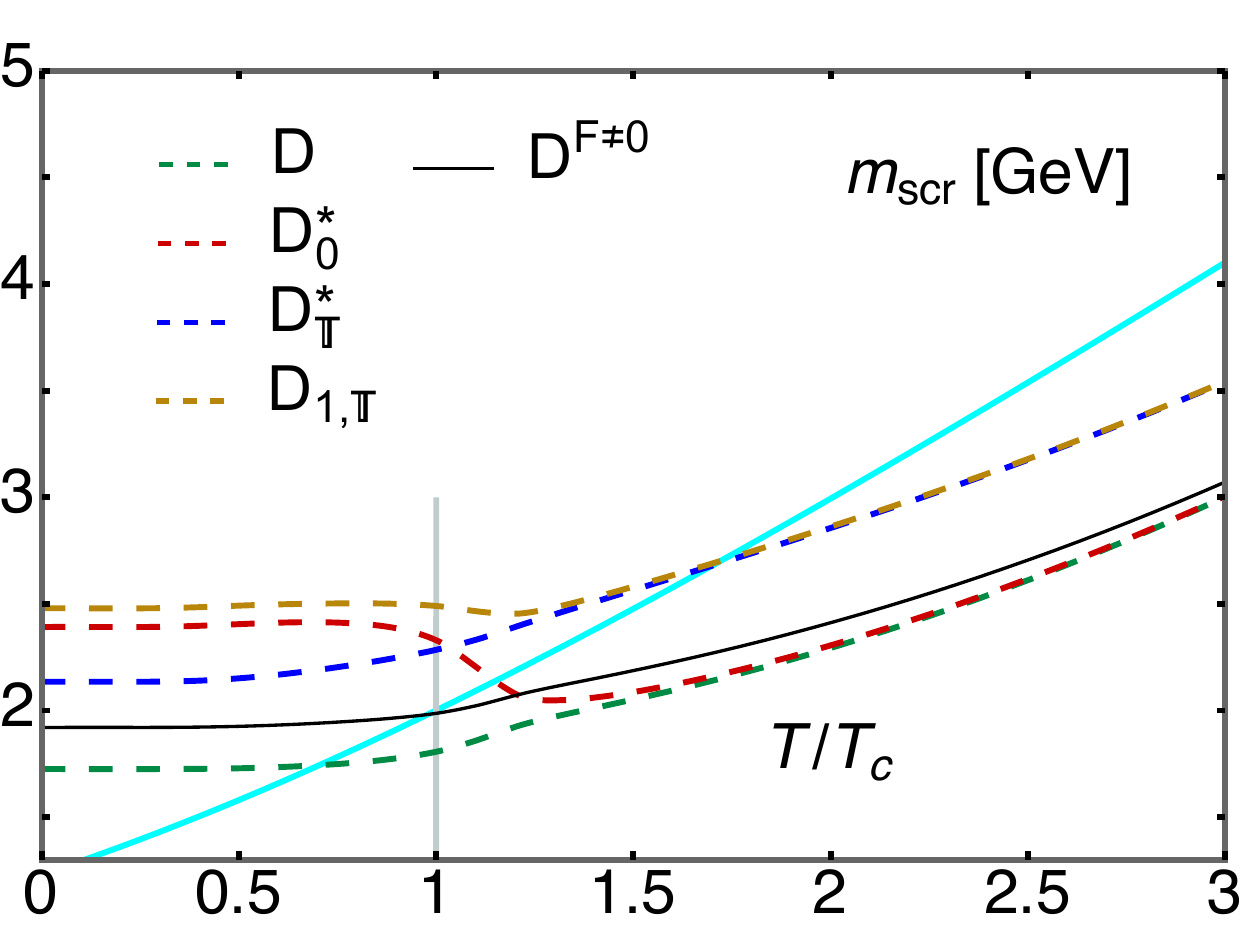}{A}%
  \hspace{\panelgap}%
  \panelplot{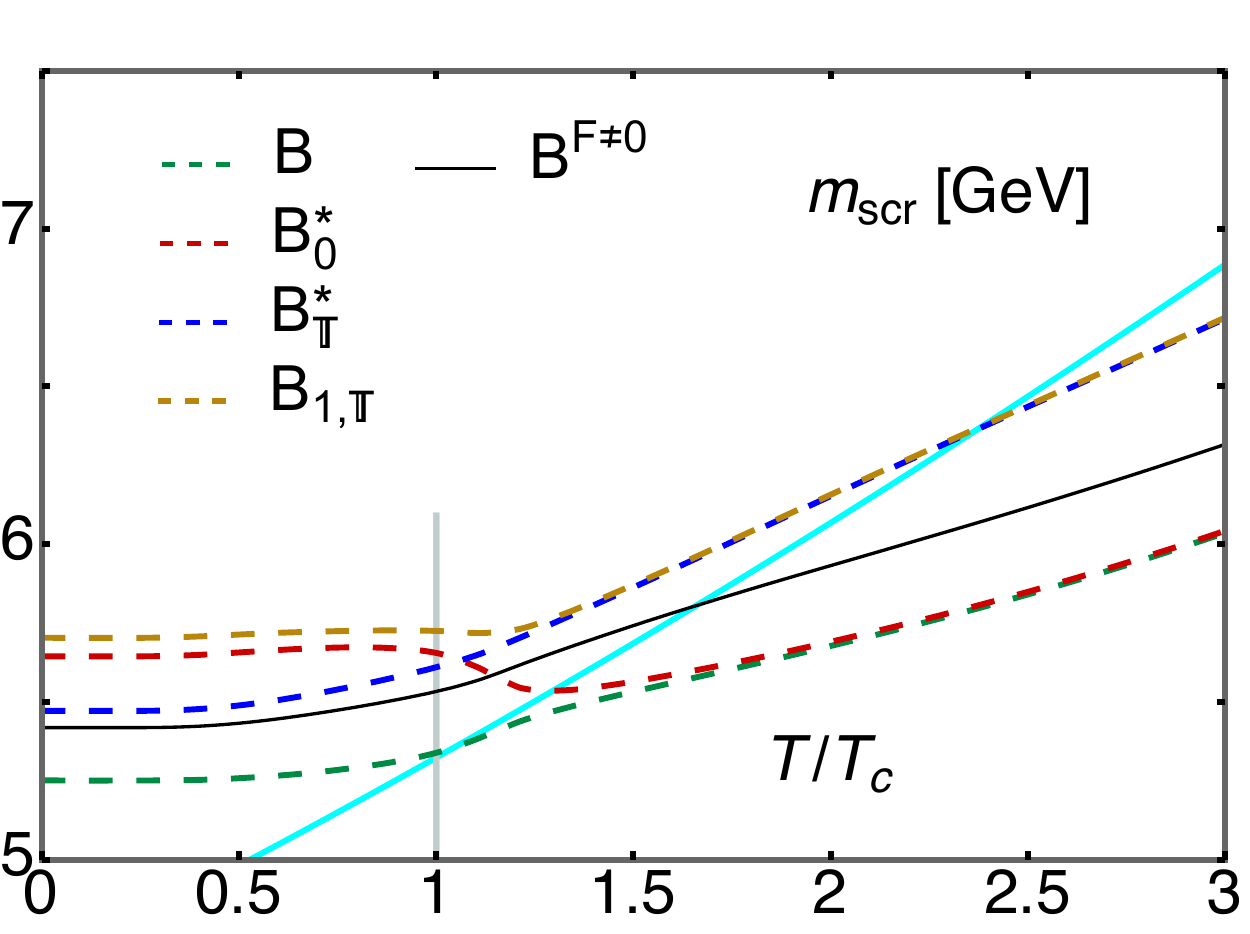}{B}%
  \hspace{\panelgap}%
  \panelplot{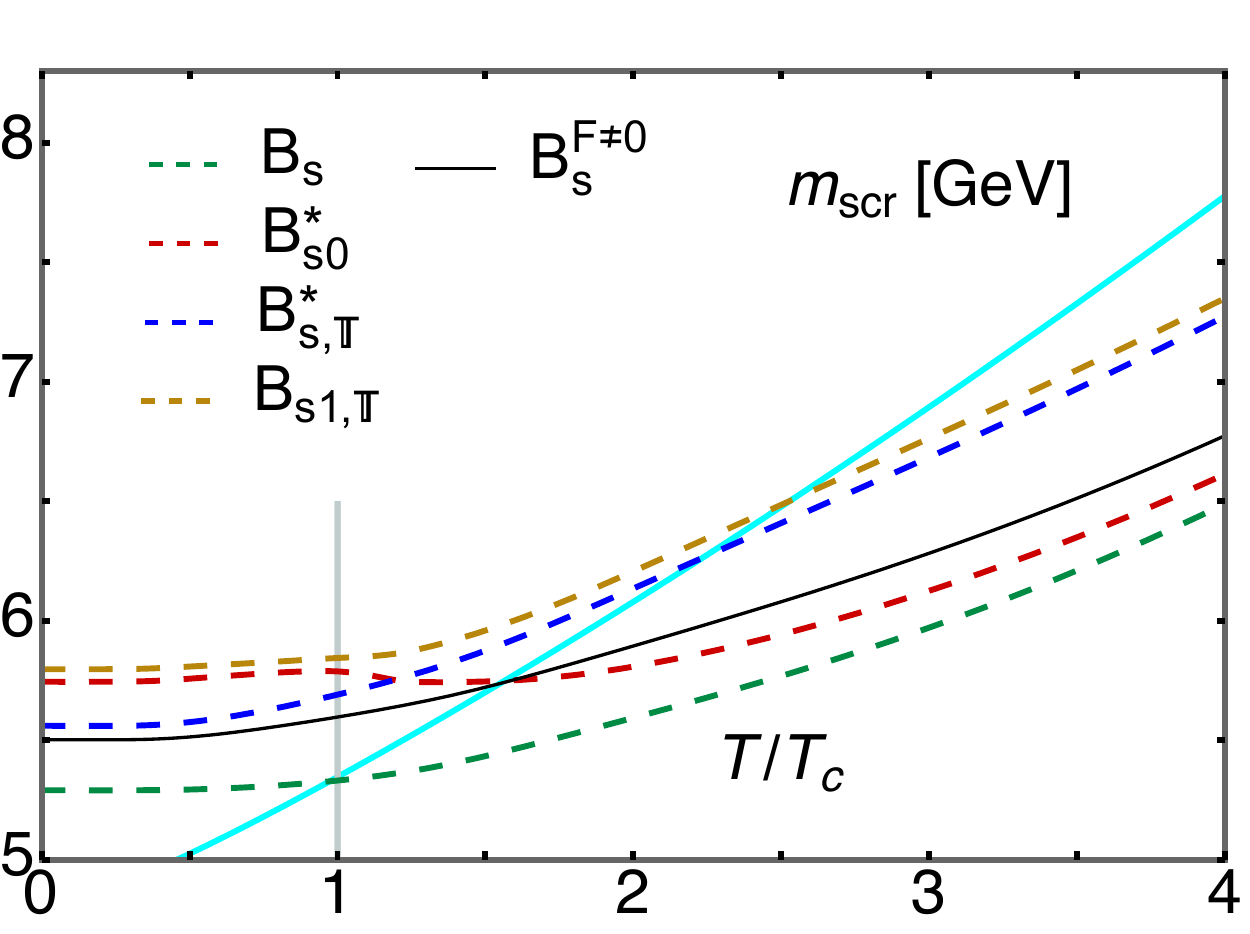}{C}%
}
\par\vspace{0.4em}
\makebox[\linewidth][c]{%
  \panelplot{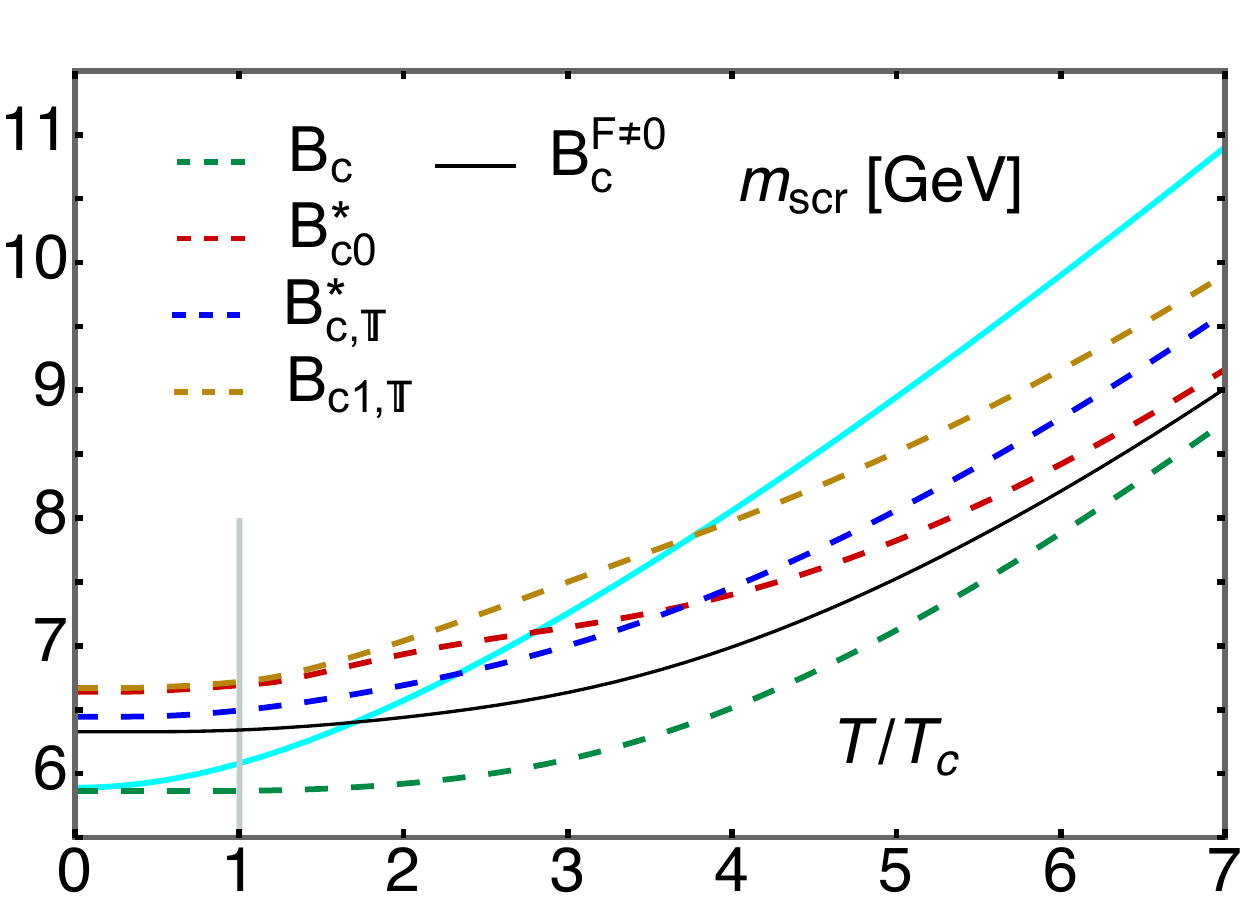}{D}%
  \hspace{\panelgap}%
  \panelplot{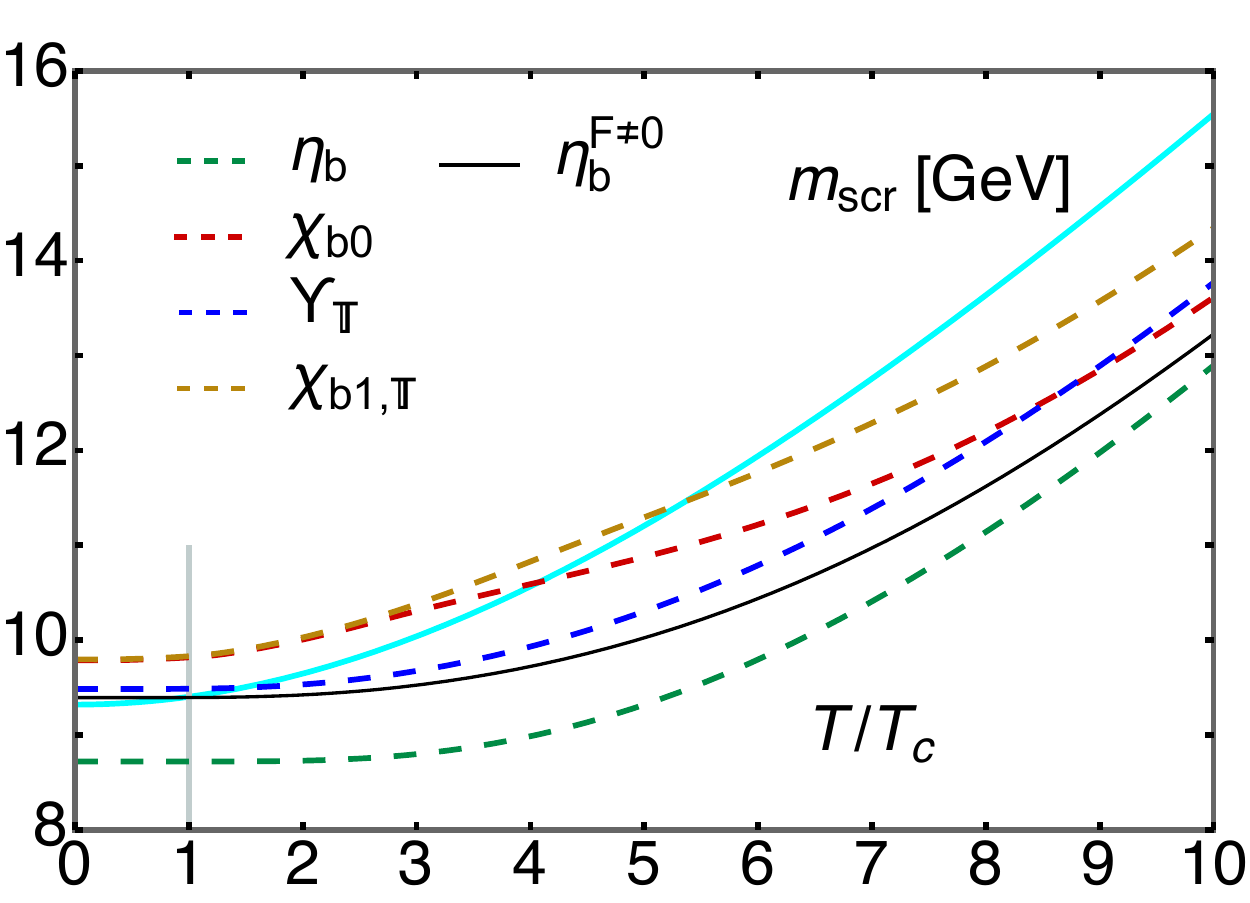}{E}%
}
\caption{\label{fig:supmscr2}
Effect of the PS pseudovector covariant on the uncorrected CI screening spectra in the prediction sectors. {\sf Panels A}--{\sf E} show the $u\bar c$, $u\bar b$, $s\bar b$, $c\bar b$, and $b\bar b$ sectors, respectively.
Curve identities are indicated in the panels, and only transverse $J=1$ modes are shown.
PS curves labeled by the superscript $F\neq0$ are the full solutions with the pseudovector covariant retained; PS curves without this superscript are the reduced $F^{0^-}\to0$ solutions used in the main text.
No medium-response correction is applied in this figure.
Cyan curves are the CI free-field limits.
This figure is the uncorrected diagnostic counterpart of Fig.\,\ref{fig:mscr2}.
}
\end{figure*}

We now quantify the effect of retaining the pseudovector covariant, the $F^{0^-}$-term, in the PS Bethe--Salpeter amplitude, Eq.\,\eqref{bsamps}, and explain why the resulting full PS solution is not used as the one-covariant PS baseline in the main text. Hereafter, \emph{full} denotes the solution with $F^{0^-}\neq0$, whereas \emph{reduced} denotes the solution obtained with $F^{0^-}\to0$. To isolate this effect, all results shown in this section are uncorrected CI screening masses; no medium-response factor is applied. For the $J=1$ channels, only the transverse modes are displayed. Figs.\,\ref{fig:supmscr1} and \ref{fig:supmscr2} are therefore diagnostic counterparts of Figs.\,\ref{fig:mscr1} and \ref{fig:mscr2}, respectively.

Fig.\,\ref{fig:supmscr1}A shows the chiral-limit result. Within numerical accuracy, the full and reduced pion screening masses coincide throughout the displayed temperature range. Hence, in the chiral limit, the pseudovector covariant does not shift the pion screening-mass eigenvalue. This behavior is consistent with the result obtained using a momentum-dependent, QCD-connected interaction in Ref.\,\cite{Maris:2000ig}.

Figs.\,\ref{fig:supmscr1}B--F and \ref{fig:supmscr2}A--E display the corresponding results for physical quark masses.
In every flavor sector, the full PS solution lies above the reduced one over the displayed temperature range.
Consequently, a direct comparison of the full PS solution with the one-covariant SC solution can suggest an artificially early approach to PS--SC degeneracy and, in some sectors, can even place the full PS screening mass above the SC mass at high temperature. This behavior is generated by the additional $F^{0^-}$ covariant and does not provide a like-for-like one-covariant baseline for ordinary PS--SC parity-partner comparisons.

This observation does not imply that the full PS solution is itself unphysical. Rather, it shows that the full two-covariant PS solution and the one-covariant SC solution are not the appropriate pair for diagnosing parity-partner convergence within the present CI construction. Accordingly, the medium-response correction is not applied to the full PS solution.
The main text uses the reduced $F^{0^-}\to0$ solution for CI-internal PS--SC comparisons, while the full solutions are retained here solely to document the effect of the pseudovector covariant.

\section{Reconstruction of the $b\bar b$ axial-vector quasi-free onset}
\label{supsec5}

This section describes how the $b\bar b$ quasi-free onset, $x^\ast_{b\bar b}$, used in the interpolation analysis is obtained from the bottomonium screening-mass results of Ref.\,\cite{Petreczky:2021zmz}. The purpose is not to define a precision lattice observable, but to obtain an operational high-temperature AX--free-field matching point in the same convention as the reference points used for the other flavor sectors.

The AX channel in Ref.\,\cite{Petreczky:2021zmz} is $h_b(1^{+-})$; that study does not provide screening data for the $\chi_{b1}(1^{++})$ channel described by our CI calculation. We therefore adopt $m_{\rm scr}^{\chi_{b1}}(T)\simeq m_{\rm scr}^{h_b}(T)$ as a phenomenological approximation for the high-temperature reconstruction and its extrapolation to the matching point. This is motivated by their nearly degenerate $1P$ vacuum masses\,\cite{ParticleDataGroup:2024cfk}, the spin-independent leading dynamics of heavy quarks, and their common leading-order free-field screening scale, Eq.\,\eqref{eq:bb_LO}.

The lattice study of Ref.\,\cite{Petreczky:2021zmz} does not tabulate the absolute $h_b$ screening mass. Instead, the relevant information is distributed over its Figs.\,1--3: Fig.\,1 shows the PS screening mass normalized by the zero-temperature $\eta_b$ mass, Fig.\,2 shows
\begin{align}
\Delta_{\rm VC-PS}(T)
=
m_{\rm scr}^{\rm VC}(T)-m_{\rm scr}^{\rm PS}(T)\,,
\end{align}
and Fig.\,3 shows
\begin{align}
\Delta_{\rm SC-PS}(T)
=
m_{\rm scr}^{\rm SC}(T)-m_{\rm scr}^{\rm PS}(T)\,.
\end{align}
The same paper states that the difference $m_{\rm scr}^{h_b}-m_{\rm scr}^{\rm VC}$ is very similar to the difference shown in Fig.\,3 and is therefore not displayed separately. With the channel identification above, we reconstruct the high-temperature AX branch as
\begin{align}
m_{\rm scr}^{\rm AX}(T)
&\simeq m_{\rm scr}^{h_b}(T)\nonumber\\
&\simeq m_{\rm scr}^{\rm PS}(T)
+\Delta_{\rm VC-PS}(T)+\Delta_{\rm SC-PS}(T)\,.
\label{eq:bb_AX_reconstruct}
\end{align}
Digitization and fit-window variations are treated as systematic uncertainties below.

Since Fig.\,1 of Ref.\,\cite{Petreczky:2021zmz} gives only the ratio $m_{\rm scr}^{\rm PS}/m_{\eta_b}^{\rm LQCD}$, we convert it to an absolute mass using the PDG value\,\cite{ParticleDataGroup:2024cfk}
\begin{align}
\label{eq:eta_b_pdg}
m_{\eta_b}^{\rm PDG}=9.40~{\rm GeV}\,.
\end{align}
Thus
\begin{subequations}
\begin{align}
\label{eq:bb_PS_abs}
m_{\rm scr}^{\rm PS}(T)
&\simeq
R_{\rm PS}(T)\,m_{\eta_b}^{\rm PDG}\,,\\
R_{\rm PS}(T)
&\equiv
\frac{m_{\rm scr}^{\rm PS}(T)}{m_{\eta_b}^{\rm LQCD}}\,.
\end{align}
\end{subequations}
This choice places the reconstructed bottomonium onset in the physical-mass convention used in the present work. It differs from the normalization used in Ref.\,\cite{Petreczky:2021zmz}, where the zero-temperature $\eta_b$ mass was estimated at the lattice bottom-quark masses.

\begin{table}[t]
\caption{\label{tab:bb_linear_coeff}
Linear parametrizations $c_0^i+c_1^iT$ used to reconstruct the high-temperature $b\bar b$ AX branch from Figs.\,1--3 of Ref.\,\cite{Petreczky:2021zmz}. The fits use the high-temperature window $0.65\lesssim T\lesssim1.0~{\rm GeV}$, with $T$ measured in GeV. The quoted coefficients correspond to one representative digitization; variations of the digitization and fit window are included in the uncertainty assigned to $x^\ast_{b\bar b}$.
}
\begin{ruledtabular}
\begin{tabular}{lccc}
 & $i$ & $c_0^i$ & $c_1^i$ \\
\hline
$R_{\rm PS}(T)$ & 1 &
0.874 & 0.253~[${\rm GeV}^{-1}$]
\\
$\Delta_{\rm VC-PS}(T)$ & 2 &
0.016~[GeV] & 0.126
\\
$\Delta_{\rm SC-PS}(T)$ & 3 &
0.268~[GeV] & $-0.072$
\end{tabular}
\end{ruledtabular}
\end{table}

Consistent with the normalization used in the main text, the leading-order (LO) free-field screening mass is
\begin{align}
\label{eq:bb_LO}
M_{\rm LO}^{b\bar b}(T)
=
2\sqrt{(\pi T)^2+(m_b^{\rm LQCD})^2}\,,
\end{align}
where $m_b^{\rm LQCD}=4.188~{\rm GeV}$ is the current mass of the bottom quark\,\cite{Petreczky:2021zmz}.

The numerical extraction proceeds as follows. We digitize the central values of $R_{\rm PS}(T)$, $\Delta_{\rm VC-PS}(T)$, and $\Delta_{\rm SC-PS}(T)$ from Figs.\,1--3 of Ref.\,\cite{Petreczky:2021zmz}. Only the high-temperature branch, $0.65\lesssim T\lesssim1.0~{\rm GeV}$, is used, since the reference point is intended to represent the parity-convergent, quasi-free screening regime rather than the rapid melting region of the $1P$ states. We fit the digitized points with the linear form
\begin{align}
\label{eq:bb_digit_fit}
c_0^i+c_1^iT\,,\quad(i=1,2,3)
\end{align}
where $T$ is measured in GeV. The coefficients of one representative digitization are listed in Table\,\ref{tab:bb_linear_coeff}. Replacing $\Delta_{\rm SC-PS}$ by a constant high-temperature average changes the final onset by less than the uncertainty quoted below.

Combining Eqs.\,\eqref{eq:bb_AX_reconstruct}--\eqref{eq:bb_digit_fit}, the reconstructed AX branch is
\begin{align}
\label{eq:bb_AX_fit}
m_{\rm scr}^{\rm AX}(T)
\simeq
m_{\eta_b}^{\rm PDG}\,[c_0^1+c_1^1T]
+
[c_0^2+c_1^2T]
+
[c_0^3+c_1^3T]\,.
\end{align}
The $b\bar b$ quasi-free onset is then defined by
\begin{align}
\label{eq:bb_anchor_eq}
m_{\rm scr}^{\rm AX}(T^\ast_{b\bar b})
=
M_{\rm LO}^{b\bar b}(T^\ast_{b\bar b})\,.
\end{align}
Using digitizations of the high-temperature points, Eq.\,\eqref{eq:bb_anchor_eq} gives
\begin{align}
\label{eq:bb_anchor_T}
T^\ast_{b\bar b}
\simeq
1.28~{\rm GeV}\,.
\end{align}
Varying the digitization, the lower end $T_{\min}$ of the fit window within $0.60$--$0.75~{\rm GeV}$, and the treatment of $\Delta_{\rm SC-PS}$ gives a spread of about $0.10$--$0.12~{\rm GeV}$. We therefore quote
\begin{align}
\label{eq:bb_anchor_T_final}
T^\ast_{b\bar b}=1.28\pm0.12~{\rm GeV}\,.
\end{align}
Using $T_c=156.5~{\rm MeV}$, this corresponds to
\begin{align}
\label{eq:bb_anchor_x_final}
x^\ast_{b\bar b}
=
\frac{T^\ast_{b\bar b}}{T_c}
=
8.18\pm0.77\,.
\end{align}
The uncertainty in Eq.\,\eqref{eq:bb_anchor_x_final} should be interpreted as a systematic onset uncertainty. It includes graphical digitization, the approximation $m_{\rm scr}^{h_b}-m_{\rm scr}^{\rm VC}\simeq m_{\rm scr}^{\rm SC}-m_{\rm scr}^{\rm PS}$, and the fit-window dependence of the high-temperature extrapolation; the PDG uncertainty in $m_{\eta_b}$ is negligible on this scale. The quoted uncertainty is conditional on the $h_b$--$\chi_{b1}$ identification and does not include an unquantified channel-dependent thermal splitting.

\begin{figure*}[t]
\centering
\includegraphics[width=0.32\textwidth]{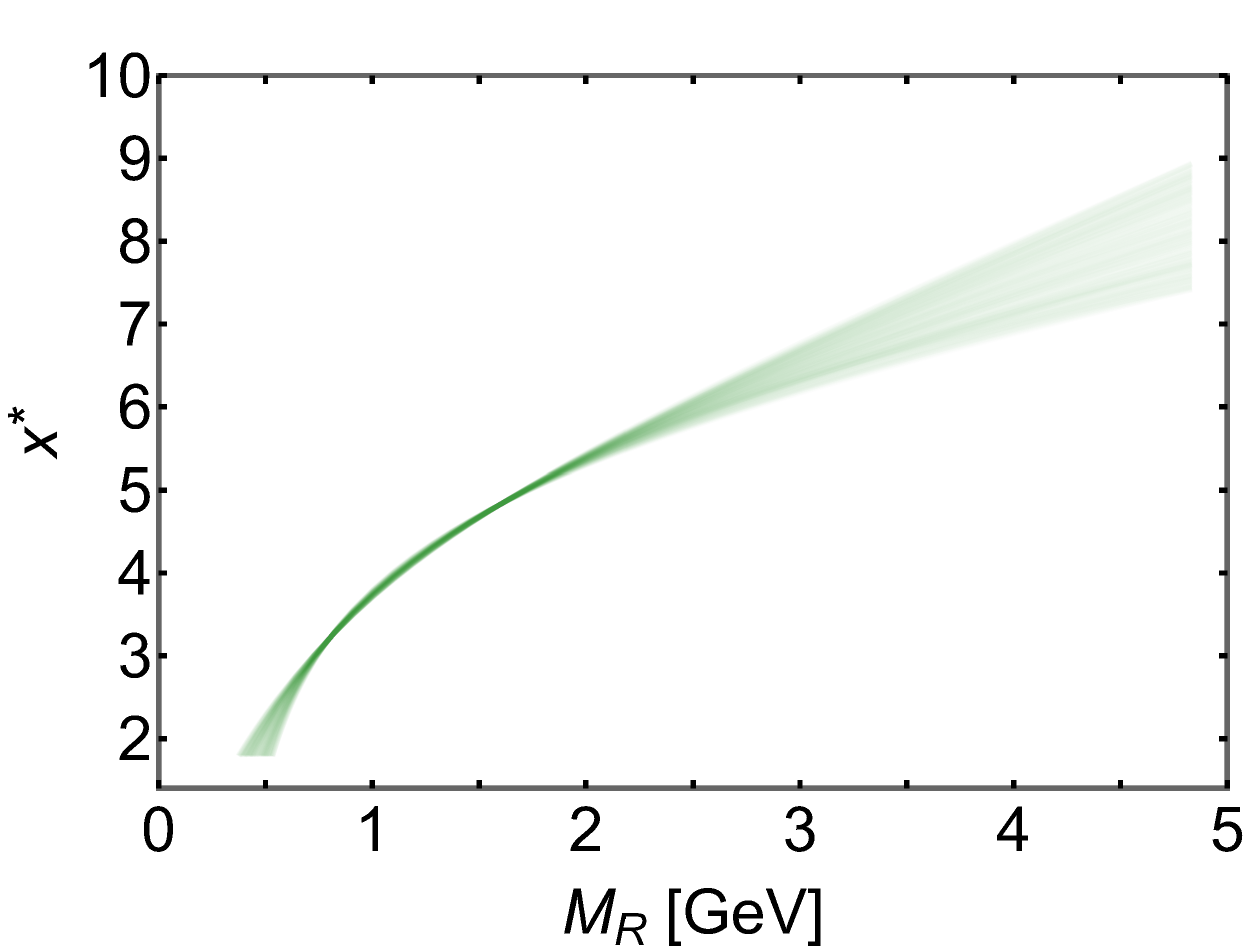}\hfill
\includegraphics[width=0.32\textwidth]{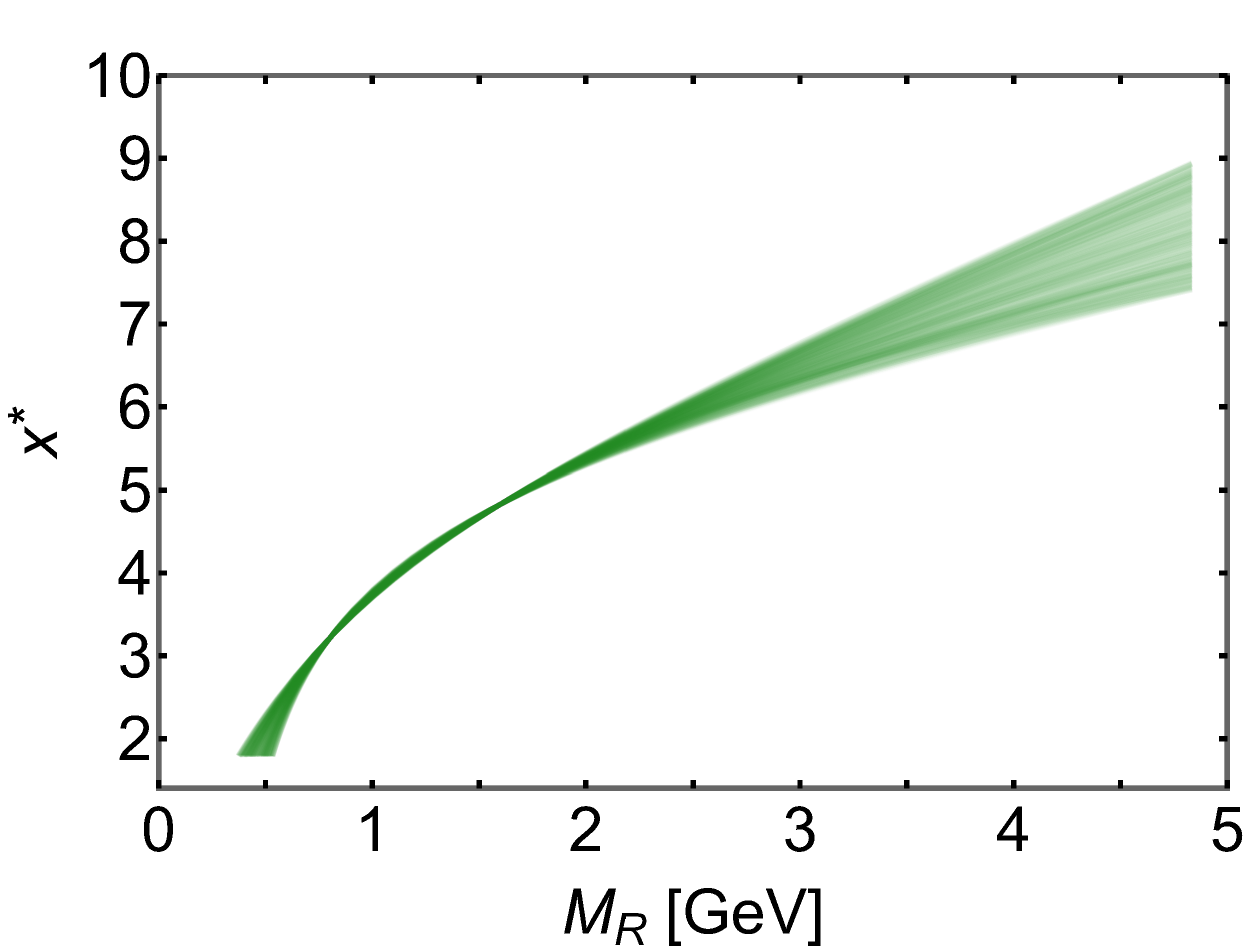}\hfill
\includegraphics[width=0.32\textwidth]{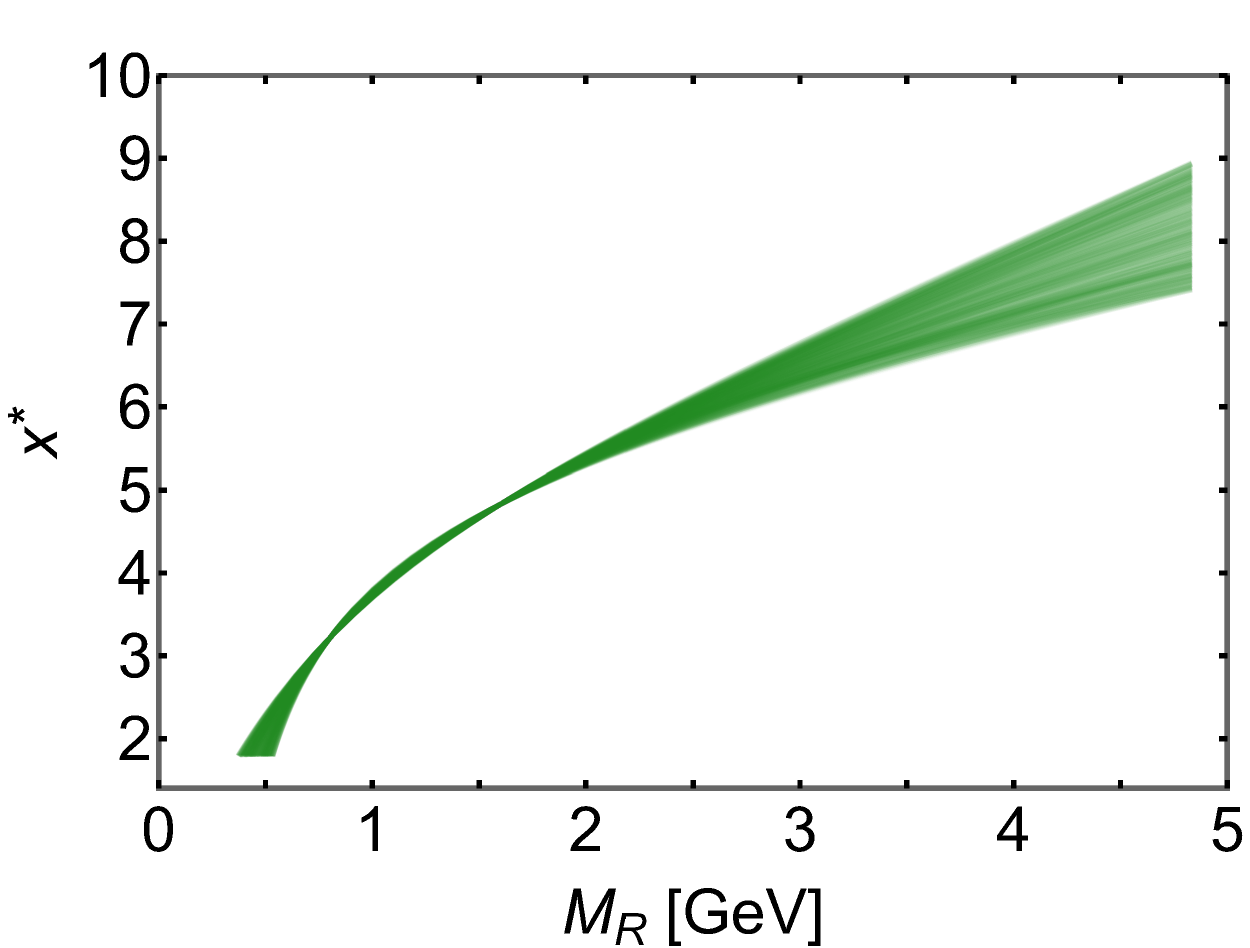}
\caption{\label{fig:spm}
Monte-Carlo SPM reconstruction of $x^\ast(M_R)$. From left to right, the panels show the envelopes obtained with $10^3$, $5\times10^3$, and $10^4$ accepted interpolants. For each interpolation, the light/strange sector is represented by a sampled point $M_R^\ell\in[M_u,M_s]$ with fixed $x^\ast_\ell=1.79$, while the bottomonium onset is sampled over $x^\ast_{b\bar b}=8.18\pm0.77$ at $M_R^{b\bar b}=M_b=4.83\,{\rm GeV}$. The $s\bar c$ and $c\bar c$ onsets are kept fixed at their central LQCD-based values. Only pole-free and monotonic interpolants in the displayed $M_R$ interval are retained. The stability of the three envelopes shows that the final $10^4$-sample envelope provides a robust systematic interpolation band for the predictions in the main text.
}
\end{figure*}


\section{Monte-Carlo Schlessinger-point interpolation of $x^\ast_{fg}$}
\label{supsec6}

We estimate the unknown quasi-free onsets $x^\ast_{fg}$ by Monte-Carlo Schlessinger point method (SPM)\,\cite{Schlessinger:1966zz,Schlessinger:1968vsk,Tripolt:2016cya,Chen:2018nsg,Cui:2020rmu,Cui:2021vgm} reconstructions in the reduced dressed-quark mass
\begin{align}
\label{eq:MR_spm}
M_R=
\frac{2M_f(0)M_g(0)}{M_f(0)+M_g(0)}\,.
\end{align}
For degenerate flavor sectors, $M_R=M_f(0)$. The purpose of the SPM analysis is to propagate the ambiguity of the common light/strange onset and the systematic uncertainty of the reconstructed $b\bar b$ onset into the predicted quasi-free onsets used in the main text.

The procedure is as follows.
\begin{description}
\item[Step 1]
For the light/strange sector, we use a single sampled onset,
\begin{align}
M_R^\ell\in[M_u,M_s]\,,
\qquad
x^\ast_\ell=1.79\,.
\end{align}
This implements the common light/strange matching prescription adopted in the main text, while allowing the associated reduced mass to span the interval from $M_u$ to $M_s$.
\item[Step 2]
For the bottomonium input, we set
\begin{align}
M_R^{b\bar b}=M_b=4.83~{\rm GeV}\,,
\end{align}
and sample
\begin{align}
x^\ast_{b\bar b}\in[8.18-0.77,\,8.18+0.77]\,,
\end{align}
using the systematic onset uncertainty obtained in Sec.\,\ref{supsec5}.
\item[Step 3]
The $s\bar c$ and $c\bar c$ onsets are kept fixed at their central
LQCD-based values:
\begin{subequations}
\begin{align}
(M_R,x^\ast)_{s\bar c}
&=
\left(
\frac{2M_sM_c}{M_s+M_c},\,3.22
\right)\,,
\\
(M_R,x^\ast)_{c\bar c}
&=
(M_c,\,4.83)\,.
\end{align}
\end{subequations}
\item[Step 4]
For each Monte-Carlo trial, four input points are used: the two fixed onsets, $s\bar c$ and $c\bar c$, and the two sampled onsets, the light/strange point and $b\bar b$. An SPM interpolant is then constructed from these four points. Only interpolants that are pole-free and monotonic over the displayed $M_R$ interval are retained.
\item[Step 5]
The construction is repeated until $10^3$, $5\times10^3$, and $10^4$ accepted interpolants are obtained. The resulting envelopes are shown in Fig.\,\ref{fig:spm}. The $10^4$-sample envelope is used as the final prediction band. This band represents a systematic interpolation uncertainty, not a statistical lattice error.
\end{description}

\begin{table}[t]
\caption{\label{tab:MRsupp}
Reduced masses and quasi-free onsets used in the main text. The first five rows are sectors with direct LQCD benchmarks. The $u\bar c$, $u\bar b$, $s\bar b$, and $c\bar b$ entries are SPM predictions. The $b\bar b$ entry is the reconstructed bottomonium input of Sec.\,\ref{supsec5}.
}
\begin{ruledtabular}
\begin{tabular}{lccc}
Flavor sector & $M_R\,[{\rm GeV}]$ & $x^\ast_{fg}$ & Status \\
\hline
$u\bar d$ & 0.37 & 1.79 & LQCD benchmark \\
$u\bar s$ & 0.44 & 1.79 & LQCD benchmark \\
$s\bar s$ & 0.53 & 1.79 & LQCD benchmark \\
$s\bar c$ & 0.80 & 3.22 & LQCD benchmark \\
$c\bar c$ & 1.60 & 4.83 & LQCD benchmark \\
$u\bar c$ & 0.60 & 2.46(18) & SPM prediction \\
$u\bar b$ & 0.69 & 2.83(8) & SPM prediction \\
$s\bar b$ & 0.96 & 3.66(5) & SPM prediction \\
$c\bar b$ & 2.40 & 5.78(17) & SPM prediction \\
$b\bar b$ & 4.83 & 8.18(77) & reconstructed input
\end{tabular}
\end{ruledtabular}
\end{table}

For reference, Table\,\ref{tab:MRsupp} collects the reduced masses used in the benchmark and prediction sectors. These values are computed from the $T=0$ dressed-quark masses in Table\,\ref{tableqmasses} using Eq.\,\eqref{eq:MR_spm}. 
In the SPM construction, the three light/strange benchmark sectors are not used as three independent interpolation inputs; instead, they are represented by the single sampled light/strange onset described in Step 1.

Using the accepted SPM ensemble, every value of $M_R\in(M_s,M_b)$, except for the fixed $s\bar c$ and $c\bar c$ inputs, yields an estimate of $x^\ast$ with a systematic uncertainty band. The resulting onsets set the upper end of the finite interval on which the medium-response function in the main text is applied in the prediction sectors.


\begin{thebibliography}{99}


\bibitem{Bazavov:2020teh}
A.~Bazavov and J.~H.~Weber,
Prog. Part. Nucl. Phys. \textbf{116}, 103823 (2021).

\bibitem{Aoki:2006we}
Y.~Aoki, G.~Endrodi, Z.~Fodor, S.~D.~Katz and K.~K.~Szabo,
Nature \textbf{443}, 675-678 (2006).

\bibitem{HotQCD:2018pds}
A.~Bazavov \textit{et al.} [HotQCD],
Phys. Lett. B \textbf{795}, 15-21 (2019).

\bibitem{Harris:2023tti}
J.~W.~Harris and B.~M{\"u}ller,
Eur. Phys. J. C \textbf{84}, no.3, 247 (2024).

\bibitem{Arslandok:2023utm}
M.~Arslandok, S.~A.~Bass, A.~A.~Baty, I.~Bautista, C.~Beattie, F.~Becattini, R.~Bellwied, Y.~Berdnikov, A.~Berdnikov and J.~Bielcik, \textit{et al.}
``Hot QCD White Paper,''
[arXiv:2303.17254 [nucl-ex]].

\bibitem{Gross:1980br}
D.~J.~Gross, R.~D.~Pisarski and L.~G.~Yaffe,
Rev. Mod. Phys. \textbf{53}, 43 (1981).

\bibitem{Braaten:1995jr}
E.~Braaten and A.~Nieto,
Phys. Rev. D \textbf{53}, 3421-3437 (1996).

\bibitem{Bala:2025ilf}
D.~Bala, O.~Kaczmarek, P.~Petreczky, S.~Sharma and S.~Tah,
Phys. Rev. Lett. \textbf{135}, no.1, 012301 (2025).

\bibitem{Kaczmarek:2022oiu}
O.~Kaczmarek,
Lect. Notes Phys. \textbf{999}, 281-305 (2022).

\bibitem{Cheng:2010fe}
M.~Cheng, S.~Datta, A.~Francis, J.~van der Heide, C.~Jung, O.~Kaczmarek, F.~Karsch, E.~Laermann, R.~D.~Mawhinney and C.~Miao, \textit{et al.}
Eur. Phys. J. C \textbf{71}, 1564 (2011).

\bibitem{Bazavov:2014cta}
A.~Bazavov, F.~Karsch, Y.~Maezawa, S.~Mukherjee and P.~Petreczky,
Phys. Rev. D \textbf{91}, no.5, 054503 (2015).

\bibitem{Bazavov:2019www}
A.~Bazavov, S.~Dentinger, H.~T.~Ding, P.~Hegde, O.~Kaczmarek, F.~Karsch, E.~Laermann, A.~Lahiri, S.~Mukherjee and H.~Ohno, \textit{et al.}
Phys. Rev. D \textbf{100}, no.9, 094510 (2019).

\bibitem{Petreczky:2021zmz}
P.~Petreczky, S.~Sharma and J.~H.~Weber,
Phys. Rev. D \textbf{104}, no.5, 054511 (2021).

\bibitem{Aoki:2025mue}
Y.~Aoki \textit{et al.} [JLQCD],
Phys. Rev. D \textbf{111}, no.11, 114506 (2025).

\bibitem{Ce:2025ihd}
M.~C{\`e}, L.~Giusti, D.~Laudicina, M.~Pepe and P.~Rescigno,
JHEP \textbf{05}, 044 (2025).

\bibitem{Laine:2003bd}
M.~Laine and M.~Vepsalainen,
JHEP \textbf{02}, 004 (2004).

\bibitem{DallaBrida:2021ddx}
M.~Dalla Brida, L.~Giusti, T.~Harris, D.~Laudicina and M.~Pepe,
JHEP \textbf{04}, 034 (2022).

\bibitem{Brandt:2014uda}
B.~B.~Brandt, A.~Francis, M.~Laine and H.~B.~Meyer,
JHEP \textbf{05}, 117 (2014).

\bibitem{CMS:2012gvv}
S.~Chatrchyan \textit{et al.} [CMS],
Phys. Rev. Lett. \textbf{109}, 222301 (2012)
[erratum: Phys. Rev. Lett. \textbf{120}, no.19, 199903 (2018)].

\bibitem{Matsui:1986dk}
T.~Matsui and H.~Satz,
Phys. Lett. B \textbf{178}, 416-422 (1986).

\bibitem{Rapp:2008tf}
R.~Rapp, D.~Blaschke and P.~Crochet,
Prog. Part. Nucl. Phys. \textbf{65}, 209-266 (2010).

\bibitem{Thews:2000rj}
R.~L.~Thews, M.~Schroedter and J.~Rafelski,
Phys. Rev. C \textbf{63}, 054905 (2001).

\bibitem{Greco:2003vf}
V.~Greco, C.~M.~Ko and R.~Rapp,
Phys. Lett. B \textbf{595}, 202-208 (2004).

\bibitem{He:2011qa}
M.~He, R.~J.~Fries and R.~Rapp,
Phys. Rev. C \textbf{86}, 014903 (2012).

\bibitem{Roberts:1994dr}
C.~D.~Roberts and A.~G.~Williams,
Prog. Part. Nucl. Phys. \textbf{33}, 477-575 (1994).

\bibitem{Roberts:2000aa}
C.~D.~Roberts and S.~M.~Schmidt,
Prog. Part. Nucl. Phys. \textbf{45}, S1-S103 (2000).

\bibitem{Eichmann:2016yit}
G.~Eichmann, H.~Sanchis-Alepuz, R.~Williams, R.~Alkofer and C.~S.~Fischer,
Prog. Part. Nucl. Phys. \textbf{91}, 1-100 (2016).

\bibitem{Fischer:2018sdj}
C.~S.~Fischer,
Prog. Part. Nucl. Phys. \textbf{105}, 1-60 (2019).

\bibitem{Maris:2000ig}
P.~Maris, C.~D.~Roberts, S.~M.~Schmidt and P.~C.~Tandy,
Phys. Rev. C \textbf{63}, 025202 (2001).

\bibitem{Blaschke:2000gd}
D.~Blaschke, G.~Burau, Y.~L.~Kalinovsky, P.~Maris and P.~C.~Tandy,
Int. J. Mod. Phys. A \textbf{16}, 2267-2291 (2001).

\bibitem{Gao:2020hwo}
F.~Gao and M.~Ding,
Eur. Phys. J. C \textbf{80}, no.12, 1171 (2020).

\bibitem{Wang:2013wk}
K.~l.~Wang, Y.~x.~Liu, L.~Chang, C.~D.~Roberts and S.~M.~Schmidt,
Phys. Rev. D \textbf{87}, no.7, 074038 (2013).

\bibitem{Chen:2024emt}
C.~Chen, F.~Gao and S.~x.~Qin,
Phys. Rev. D \textbf{112}, no.1, 014022 (2025).

\bibitem{Ramirez-Garrido:2025rsu}
M.~A.~Ram{\'\i}rez-Garrido, R.~J.~Hern{\'a}ndez-Pinto, I.~M.~Higuera-Angulo and L.~X.~Guti{\'e}rrez-Guerrero,
Phys. Rev. D \textbf{112}, no.3, 034024 (2025).

\bibitem{Gutierrez-Guerrero:2026hhu}
L.~X.~Guti{\'e}rrez-Guerrero, M.~A.~Ram{\'\i}rez-Garrido, M.~A.~P.~de Le{\'o}n and R.~J.~Hern{\'a}ndez-Pinto,
Phys. Rev. D \textbf{113}, no.1, 014035 (2026).

\bibitem{Schlessinger:1966zz}
L.~Schlessinger and C.~Schwartz,
Phys. Rev. Lett. \textbf{16}, 1173-1174 (1966).

\bibitem{Schlessinger:1968vsk}
L.~Schlessinger,
Phys. Rev. \textbf{167}, no.5, 1411 (1968).

\bibitem{Tripolt:2016cya}
R.~A.~Tripolt, I.~Haritan, J.~Wambach and N.~Moiseyev,
Phys. Lett. B \textbf{774}, 411-416 (2017).

\bibitem{Chen:2018nsg}
C.~Chen, Y.~Lu, D.~Binosi, C.~D.~Roberts, J.~Rodr{\'\i}guez-Quintero and J.~Segovia,
Phys. Rev. D \textbf{99}, no.3, 034013 (2019).

\bibitem{Cui:2020rmu}
Z.~F.~Cui, C.~Chen, D.~Binosi, F.~de Soto, C.~D.~Roberts, J.~Rodr{\'\i}guez-Quintero, S.~M.~Schmidt and J.~Segovia,
Phys. Rev. D \textbf{102}, no.1, 014043 (2020).

\bibitem{Cui:2021vgm}
Z.~F.~Cui, D.~Binosi, C.~D.~Roberts and S.~M.~Schmidt,
Phys. Rev. Lett. \textbf{127}, no.9, 092001 (2021).

\bibitem{Chen:2012qr}
C.~Chen, L.~Chang, C.~D.~Roberts, S.~Wan and D.~J.~Wilson,
Few Body Syst. \textbf{53}, 293-326 (2012).

\bibitem{Munczek:1994zz}
H.~J.~Munczek,
Phys. Rev. D \textbf{52}, 4736-4740 (1995).

\bibitem{Bender:1996bb}
A.~Bender, C.~D.~Roberts and L.~Von Smekal,
Phys. Lett. B \textbf{380}, 7-12 (1996).

\bibitem{Chang:2009zb}
L.~Chang and C.~D.~Roberts,
Phys. Rev. Lett. \textbf{103}, 081601 (2009).

\bibitem{Fischer:2009jm}
C.~S.~Fischer and R.~Williams,
Phys. Rev. Lett. \textbf{103}, 122001 (2009).

\bibitem{Williams:2015cvx}
R.~Williams, C.~S.~Fischer and W.~Heupel,
Phys. Rev. D \textbf{93}, no.3, 034026 (2016).

\bibitem{Qin:2020jig}
S.~X.~Qin and C.~D.~Roberts,
Chin. Phys. Lett. \textbf{38}, no.7, 071201 (2021).

\bibitem{Bhagwat:2007ha}
M.~S.~Bhagwat, L.~Chang, Y.~X.~Liu, C.~D.~Roberts and P.~C.~Tandy,
Phys. Rev. C \textbf{76}, 045203 (2007).

\bibitem{Ding:2018xwy}
M.~Ding, K.~Raya, A.~Bashir, D.~Binosi, L.~Chang, M.~Chen and C.~D.~Roberts,
Phys. Rev. D \textbf{99}, no.1, 014014 (2019).

\bibitem{Binosi:2016nme}
D.~Binosi, C.~Mezrag, J.~Papavassiliou, C.~D.~Roberts and J.~Rodriguez-Quintero,
Phys. Rev. D \textbf{96}, no.5, 054026 (2017).

\bibitem{Ebert:1996vx}
D.~Ebert, T.~Feldmann and H.~Reinhardt,
Phys. Lett. B \textbf{388}, 154-160 (1996).

\bibitem{Krein:1990sf}
G.~Krein, C.~D.~Roberts and A.~G.~Williams,
Int. J. Mod. Phys. A \textbf{7}, 5607-5624 (1992).

\bibitem{Yin:2021uom}
P.~L.~Yin, Z.~F.~Cui, C.~D.~Roberts and J.~Segovia,
Eur. Phys. J. C \textbf{81}, no.4, 327 (2021).

\bibitem{Bazavov:2011nk}
A.~Bazavov, T.~Bhattacharya, M.~Cheng, C.~DeTar, H.~T.~Ding, S.~Gottlieb, R.~Gupta, P.~Hegde, U.~M.~Heller and F.~Karsch, \textit{et al.}
Phys. Rev. D \textbf{85}, 054503 (2012).

\bibitem{Fukushima:2013rx}
K.~Fukushima and C.~Sasaki,
Prog. Part. Nucl. Phys. \textbf{72}, 99-154 (2013).

\bibitem{Buballa:2003qv}
M.~Buballa,
Phys. Rept. \textbf{407}, 205-376 (2005).

\bibitem{Lu:2017cln}
Y.~Lu, C.~Chen, C.~D.~Roberts, J.~Segovia, S.~S.~Xu and H.~S.~Zong,
Phys. Rev. C \textbf{96}, no.1, 015208 (2017).

\bibitem{Yin:2019bxe}
P.~L.~Yin, C.~Chen, G.~Krein, C.~D.~Roberts, J.~Segovia and S.~S.~Xu,
Phys. Rev. D \textbf{100}, no.3, 034008 (2019).

\bibitem{Cheng:2022jxe}
P.~Cheng, F.~E.~Serna, Z.~Q.~Yao, C.~Chen, Z.~F.~Cui and C.~D.~Roberts,
Phys. Rev. D \textbf{106}, no.5, 054031 (2022).

\bibitem{ParticleDataGroup:2024cfk}
S.~Navas \textit{et al.} [Particle Data Group],
Phys. Rev. D \textbf{110}, no.3, 030001 (2024).

\bibitem{Mathur:2018epb}
N.~Mathur, M.~Padmanath and S.~Mondal,
Phys. Rev. Lett. \textbf{121}, no.20, 202002 (2018).

\bibitem{Dowdall:2013rya}
R.~J.~Dowdall, C.~T.~H.~Davies, G.~P.~Lepage and C.~McNeile,
Phys. Rev. D \textbf{88}, 074504 (2013).







\end{thebibliography}
\end{document}